\documentclass[showpacs,aps,prd,nofootinbib,floatfix,amsmath,amssymb,twocolumn]{revtex4}
\usepackage{mathrsfs}
\usepackage{graphicx}
\usepackage[dvipsnames]{xcolor}
\usepackage{dsfont}
\begin{document}

\makeatletter
\newbox\slashbox \setbox\slashbox=\hbox{$/$}
\newbox\Slashbox \setbox\Slashbox=\hbox{\large$/$}
\def\pFMslash#1{\setbox\@tempboxa=\hbox{$#1$}
  \@tempdima=0.5\wd\slashbox \advance\@tempdima 0.5\wd\@tempboxa
  \copy\slashbox \kern-\@tempdima \box\@tempboxa}
\def\pFMSlash#1{\setbox\@tempboxa=\hbox{$#1$}
  \@tempdima=0.5\wd\Slashbox \advance\@tempdima 0.5\wd\@tempboxa
  \copy\Slashbox \kern-\@tempdima \box\@tempboxa}
\def\FMslash{\protect\pFMslash}
\def\FMSlash{\protect\pFMSlash}
\def\miss#1{\ifmmode{/\mkern-11mu #1}\else{${/\mkern-11mu #1}$}\fi}
\makeatother

\title{One-loop contributions to $WWZ$ from a seesaw variant with radiatively-induced light-neutrino masses}

\author{H\'ector Novales-S\'anchez$^{(a)}$}\author{M\'onica Salinas$^{(b)}$}\author{Humberto V\'azquez-Castro$^{(a)}$}
\affiliation{
$^{(a)}$Facultad de Ciencias F\'isico Matem\'aticas, Benem\'erita Universidad Aut\'onoma de Puebla, Apartado Postal 1152 Puebla, Puebla, M\'exico\\$^{(b)}$Departamento de F\'isica, Centro de Investigaci\'on y de Estudios Avanzados del IPN, Apartado Postal 14-740,07000 Ciudad de M\'exico, M\'exico}

\begin{abstract}
The experimental confirmation of neutrino mass and mixing disagrees with the picture of the Standard Model, and thus provides a good motivation to look for unknown physics. A seesaw variant in which neutrino masses are radiatively generated, with neutrinos characterized by Majorana fields and where a set of three hypothetical neutral heavy leptons are present, has been considered in the present work. We have calculated, estimated, and analyzed the contributions from virtual light and heavy neutrinos to the gauge vertex $WWZ$, in the context of a future electron-positron collider. We have found that contributions to both $CP$-even and $CP$-odd anomalous couplings are generated. Our estimations indicate that contributions as large as $~10^{-3}$ can be achieved in both cases. This is particularly interesting for the $CP$-odd effects, which are expected to appear in the framework of the Standard Model since the three-loop order, thus being quite suppressed. By considering the expected sensitivity of the International Linear Collider to this gauge vertex, we conclude that these new-physics effects would be within its reach for $e^+e^-$ collisions taking place at a center-of-mass energy of $800\,{\rm GeV}$.
\end{abstract}


\maketitle

\section{Introduction}
The evidence confirming neutrino oscillations, first achieved by experimental collaborations at the Super-Kamiokande experiment~\cite{Kamiokande} and at the Sudbury Neutrino Observatory~\cite{SNO}, showed that the neutrino sector of the Standard Model~\cite{Glashow,Salam,Weinberg} (SM) does not provide an accurate description of nature and thus requires to be extended. Massiveness of neutrinos, entailed by the phenomenon of neutrino oscillations~\cite{Pontecorvo}, might be introduced by assuming neutrinos to be characterized by Dirac fields~\cite{Dirac} and then endowing them with masses just the way it is done with all other fermions in the Yukawa sector of the SM. However, aiming at a more sort of natural definition of neutrino masses, in the sense of explaining their conspicuous tininess, alternative mechanisms have to be searched for. A well-known neutrino-mass generating mechanism is the seesaw~\cite{MoSe1,MoSe2,PaSa}, which propounds that neutrinos, assumed to be described by Majorana fermion fields~\cite{Majorana}, have such small masses due to some high-energy scale, $\Lambda$, associated to a yet-unknown physical description, beyond the Standard Model. The Weinberg operator~\cite{Weinbergoperator}, an effective-Lagrangian term~\cite{DGMP} with units $\big( {\rm mass} \big)^5$ and which introduces violation of lepton number, generates, driven by the Brout-Englert-Higgs mechanism~\cite{EnBr,Higgs}, Majorana-mass terms for neutrinos, with masses given by $m_{\nu_j}\sim\frac{v^2}{\Lambda}$, where $v=246\,{\rm GeV}$ is the SM Higgs vacuum expectation value. This mass profile matches the one emerged from the seesaw mechanism, which, in addition, comes along with a set of heavy-neutral leptons, dubbed ``heavy-neutrinos'', with masses $m_{N_j}\sim\Lambda$. In this context, current upper bounds on neutrino mass~\cite{KATRIN,SDSSexperiment,Planckexperiment}, lying within the sub-eV regime, push the energy scale $\Lambda$ towards enormous values, thus yielding huge heavy-neutrino masses. Therefore, direct production of heavy neutrinos or measurement of their quantum effects on SM observables do not seem to be achievable. So, while from a purely theoretical viewpoint, the seesaw explanation is quite appealing, it comes along with a practical drawback: the very large size of the high-energy scale $\Lambda$ avoids any possibility of measuring the new physics (NP) through current and future experimental facilities, perhaps even in the long term.  Pursuing a seesaw-like neutrino-mass origin in which heavy neutrinos bear masses with a more reasonable size, seesaw variants have been conceived\footnote{For a review on seesaw variants, see Ref.~\cite{CHLR} and references therein.}. The inverse seesaw~\cite{MoVa,GoVa,DeVa} and the linear seesaw~\cite{ALSV1,ALSV2} neutrino-mass-generating mechanisms are well-known instances. Among the available seesaw-type mechanisms, we have considered the model given in Ref.~\cite{Pilaftsis} as the framework for the present investigation. The author of this reference introduced, in the context of the seesaw mechanism, a condition to render all tree-level light-neutrino mass-terms zero, while keeping masses of heavy neutrinos untouched. This breaks the seesaw-type link among heavy- and light-neutrino masses, which allows for quite smaller masses for the heavy neutral leptons. Then, masses of light neutrinos are rather generated by radiative corrections, which defines a new connection of neutrino masses, namely, in order for light-neutrinos to have adequate tiny masses, the set of heavy-neutrino masses has to be quasi-degenerate. 
\\

\begin{figure}[ht]
\center
\includegraphics[width=7cm]{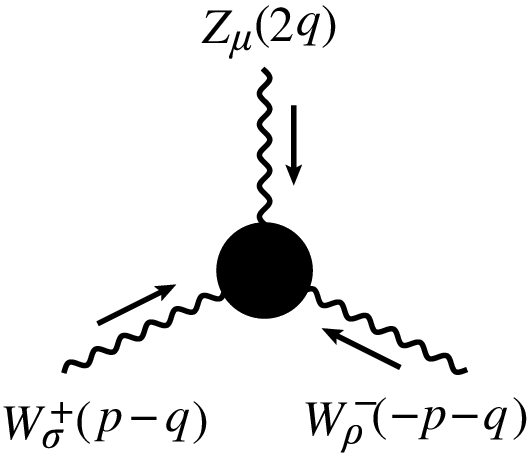}
\caption{\label{WWZ} Conventions for the $WWZ$ vertex, as first posed in Ref.~\cite{BGL}. Throughout the present paper, the external $Z$ boson is assumed to be off shell, whereas the external $W$ bosons are taken on shell.}
\end{figure}

In accordance with the technique of Feynman diagrams~\cite{Feynman}, physics beyond the SM might yield modifications on low-energy observables through diagrams in which virtual lines associated to heavy dynamic variables from some NP description participate. This is the case of the triple gauge vertex $WWZ$, which is parametrized by a vertex function
\begin{equation}
\Gamma^{WWZ}_{\sigma\rho\mu}=\Gamma_{\sigma\rho\mu}^{\rm even}+\Gamma_{\sigma\rho\mu}^{\rm odd},
\end{equation}
where $\Gamma_{\sigma\rho\mu}^{\rm even}$ is associated to $CP$-conserving effects, whereas $\Gamma_{\sigma\rho\mu}^{\rm odd}$ characterizes physics not preserving $CP$ symmetry. If the $W$-boson external lines are taken on shell, while assuming that the external $Z$-boson is off shell, the $CP$-even and $CP$-odd terms, $\Gamma_{\sigma\rho\mu}^{\rm even}$ and $\Gamma_{\sigma\rho\mu}^{\rm odd}$, are respectively given by\footnote{A brief discussion on the effective-Lagrangian origin of both vertex functions $\Gamma^{\rm even}_{\sigma\rho\mu}$ and $\Gamma^{\rm odd}_{\sigma\rho\mu}$ is provided in Appendix~\ref{App2}}~\cite{BGL,HPZH,BaZe}
\begin{eqnarray}
&&
\Gamma^{\rm even}_{\sigma\rho\mu}=ig_Z\Big( g_1\big( 2p_\mu g_{\sigma\rho} +4( q_\rho g_{\sigma\mu}-q_\sigma g_{\rho\mu} ) \big)
\nonumber \\
&&
\hspace{1.1cm}
+\frac{4\Delta Q}{m_W^2}p_\mu\Big( q_\sigma q_\rho-\frac{q^2}{2}g_{\sigma\rho} \Big)
\nonumber \\
&&
\hspace{1.1cm}
+2\Delta\kappa(q_\rho g_{\sigma\mu}-q_\sigma g_{\rho\mu})+if_1\,\epsilon_{\sigma\rho\mu\alpha}p^\alpha
\Big),
\label{Gammaeven}
\end{eqnarray}
\begin{eqnarray}
&&
\Gamma^{\rm odd}_{\sigma\rho\mu}=ig_Z\Big(
2\Delta\tilde{\kappa}\epsilon_{\sigma\rho\mu\alpha}q^\alpha+\frac{4\Delta\tilde{Q}}{m_W^2}q_\rho\epsilon_{\sigma\mu\alpha\beta} p^\alpha q^\beta
\nonumber \\
&&\hspace{1cm}
+i\tilde{f}_1(q_\rho g_{\sigma\mu}+q_\sigma g_{\rho\mu})
\nonumber \\
&&\hspace{1cm}
+\tilde{f}_2\,p^\lambda\epsilon_{\sigma\rho\lambda\alpha}(q^2\delta^\alpha\hspace{0.0001cm}_\mu-q^\alpha q_\mu)
\Big),
\label{Gammaodd}
\end{eqnarray}
in accordance with the conventions of Fig.~\ref{WWZ}.
In this equation, $g_Z=e\cot\theta_{\rm W}$, where $\theta_{\rm W}$ is the weak mixing angle and $e$ denotes the electric charge of a positron. 
In the SM, $g_1=1$ at the tree level, whereas all other form factors $\Delta Q$, $\Delta\kappa$, $f_1$, $\Delta\tilde\kappa$, $\Delta\tilde Q$, $\tilde{f}_1$, and $\tilde{f}_2$ vanish, so they are commonly referred to by ``anomalous couplings'' (AC). The one-loop SM contributions to $\Delta\kappa$, $\Delta Q$, $\Delta\tilde{\kappa}$, and $\Delta\tilde{Q}$ have been already calculated~\cite{AKLPS,PaPhi}, from which $CP$-conserving contributions of orders $\Delta\kappa_{\rm SM}\sim10^{-3}$ and $\Delta Q_{\rm SM}\sim10^{-4}$ have been estimated. On the other hand, $CP$-violating contributions from the SM are absent at the one-loop level, though their emergence at the level of three loops is expected~\cite{CKL,PoKhr}. 
\\

Our interest, in the present investigation, focuses on the NP contributions from virtual Majorana neutrinos to $WWZ$, produced at one loop, for which we consider both light and heavy neutrinos, in the framework defined by the model of Ref.~\cite{Pilaftsis}. The associated NP contributions to the AC $\Delta\kappa$, $\Delta Q$, $\Delta\tilde\kappa$, and $\Delta\tilde Q$, characterizing this gauge vertex, are computed, estimated and analyzed. 
While both $CP$-conserving contributions, $\Delta Q$ and $\Delta\kappa$, are generated, the $CP$-odd $\Delta\tilde{Q}$ is found to vanish, so the only nonzero $CP$-violating contribution turns out to be $\Delta\tilde{\kappa}$. Anticipating the presumable presence of ultraviolet divergences, we utilize the dimensional-regularization method~\cite{BoGi,tHooVe}. At the end of the day, we arrive at the conclusion that all the generated contributions to $\Delta Q$, $\Delta\kappa$, and $\Delta\tilde{\kappa}$ are ultraviolet finite. Feynman rules for Dirac fields differ from those for Majorana fermions~\cite{DEHK,GlZr}. In particular, given some physical process, the Majorana framework usually involves a larger number of contributing diagrams than the Dirac case. In the present work, the set of Feynman diagrams contributing via virtual Majorana fermions includes (1) those diagrams which would contribute to the same process if such fermions were assumed to be of Dirac type, and (2) a set of extra diagrams to be taken into account. We call the extra diagrams, characteristic of Majorana fermions, ``Majorana-type diagrams'', whereas those which occur in both the Dirac and Majorana cases are referred to as ``Dirac-type diagrams''. We find that the contributions from Majorana-type diagrams exactly match those of Dirac-type diagrams. Our numerical analysis of the $WWZ$ vertex is carried out in the context of some future electron-positron colliding facility, such as the highly anticipated International Linear Collider (ILC)~\cite{ILC1,ILC2}. We consider the process $e^+e^-\to W^+W^-$, which receives contributions from an $s$-channel diagram involving a virtual $Z$-boson line. Therefore, our calculation is performed under the assumption that the external $Z$-boson line is off the mass shell, in which case, in accordance with the notation defined by Fig.~\ref{WWZ}, the $WWZ$ vertex function $\Gamma^{WWZ}_{\sigma\rho\mu}$ depends on the squared 4-momentum $(2q)^2$. Moreover, the center-of-mass energy (CME), $\sqrt{s}$, of the corresponding $e^+e^-$ collision relates to this momentum as $s=4q^2$, so all contributions to AC are CME dependent. The present investigation aims at the calculation of such off-shell contributions, as well as its analysis and estimation, in the light of the expected ILC sensitivity~\cite{ILC2}. Our estimations indicate that $CP$-preserving contributions $\Delta\kappa$ as large as $\sim10^{-3}$ can be generated for an electron-positron collision at a CME of $800\,{\rm GeV}$, thus falling into the sensitivity region expected for the ILC. On the other hand, the only nonzero $CP$-odd contribution, $\Delta\tilde\kappa$, also reaches $\sim10^{-3}$, which lies about one order of magnitude out of the reach of ILC expected sensitivity.
\\

The rest of the paper has been organized in the following manner: our analytical calculation of the one-loop contributions from Majorana neutrinos to $WWZ$, carried out in accordance with the theoretical framework provided in Appendices~\ref{App2} and \ref{App1}, is discussed in Section~\ref{AnalyticPheno}; then, an estimation of contributions is performed in Section~\ref{NumPheno}, where a discussion on our results is developed as well; and, finally, we present a summary in Section~\ref{Concs}.
\\


\section{Virtual Majorana neutrinos contribution to $WWZ$ at one loop}
\label{AnalyticPheno}
The main purpose of this section is the description and discussion of our calculation of the contributions from the neutrino model of Ref.~\cite{Pilaftsis} to the $WWZ$ vertex function. Details on the neutrino-mass model, relevant to this calculation, have been comprehensively discussed in Refs.~\cite{Pilaftsis,MMNS,NoSa}, so this information is briefly addressed in Appendix~\ref{App1}.
\\

Since the $WWZ$ vertex is assumed to be a contributing piece of an $s$-channel diagram for $e^+e^-\to W^+W^-$, the $Z$-boson external line is taken off shell, whereas both external $W$-boson lines are considered to be on the mass shell. As the calculation is carried out at the one-loop level, the contributions come exclusively from fermion loops, in which case all virtual lines are associated to fermion fields. Therefore, even though the vertex is off shell, no issues associated to gauge dependence arise. While the vertices $WWZ$ and $WW\gamma$ share some features, they have important differences. For starters, from the viewpoint of the Feynman-diagrams technique, a difference emerges, namely, fermion-loop contributions to $WWZ$ come along with extra contributing diagrams, with respect to those contributing to $WW\gamma$, which is due to the nonzero coupling $Z\nu\nu$, not occurring in the electromagnetic case. Nonetheless, the main discrepancy is, perhaps, the requirement of ${\rm U}(1)_e$ gauge invariance, which imposes restrictions on the occurrence of $WW\gamma$ couplings, but which is inoffensive to $WWZ$. If the $Z$-boson field $Z_\mu$ is replaced by the photon field $A_\mu$ in Eqs.~(\ref{LCPeven}) and (\ref{LCPodd}), electromagnetic gauge invariance forbids the generation of couplings $g_2$ and $\tilde{g}_1$, which, in contraposition, can be present in the case of $WWZ$. From the perspective of the ${\rm SU}(2)_L\otimes{\rm U}(1)_e$-invariant effective Lagrangian, which extends the electroweak SM~\cite{LLR,BuWy,Wudka}, such couplings are generated by Lagrangian terms with mass units $>({\rm mass})^6$~\cite{EllWu}, thus being subjected to a high-energy scale suppression $\Lambda^{-n}$, with $n>2$. So, due to electromagnetic gauge symmetry, the $WW\gamma$ vertex function emerged from such a Lagrangian fulfills the Ward identity~\cite{Ward}, whereas the $WWZ$ vertex function is not required to abide by it. Note, however, that invariance under Becchi-Rouet-Stora-Tyutin transformations~\cite{BRS1,BRS2,Tyutin}, a local symmetry which remains at the quantum level after the gauge has been necessarily fixed\footnote{For a review on the BRST symmetry at both the classical and quantum levels, in the context of the field-antifield formalism, see Ref.~\cite{GPS}.}, induces Slavnov-Taylor identities~\cite{Taylor,Slavnov,BSH}, which relate Green's functions. In this context, the $WWZ$ vertex function is expected participate in some alike relation. Slavnov-Taylor identities have an intricate structure, as compared to Ward identities, and include ghost-field contributions. To this respect, we find worth mentioning that the Background Field Method~\cite{DeWitt,tHooftBFM,Abbott1,Abbott2}, an unconventional quantization method aimed at the preservation of gauge invariance at the quantum level, and the Pinch Technique~\cite{Cornwall,CoPa,Papavassiliou}, a diagrammatic method meant to get well-behaved, gauge invariant, and gauge independent Green's functions, have been found to yield simple Ward-like identities for the $WWZ$ vertex function~\cite{DDW,PapavassiliouRef}.
\\

A main difference among the Dirac and the Majorana descriptions of fermions is that in the case of Majorana fields the relation $\psi^{\rm c}=\psi$, with $\psi^{\rm c}=C\overline{\psi}^{\rm T}$ the charge-conjugated field of $\psi$ and $C$ the charge-conjugation matrix, holds. This equation, known as the Majorana condition, can be fulfilled only if $\psi$ is electrically neutral. The so-far observed electromagnetic neutrality of the neutrinos~\cite{PDG} and the relatively recent confirmation, supported by neutrino oscillations, that neutrinos are massive, has fed the interest in the possibility that neutrinos are Majorana fermions~\cite{Majorana}, unlike the rest of the SM fermion particles, which abide by the Dirac description~\cite{Dirac}. Whether the description of neutrinos is correctly achieved by Dirac or Majorana fields is a question whose answer incarnates a main objective in neutrino physics. The most representative physical process to look for Majorana neutrinos is the neutrinoless double beta decay, whose observation would be evidence of the Majorana nature of the neutrinos. A plethora of experimental collaborations have been working for several years, looking for a measurement of this physical process~\cite{CUPIDMOndbd,CUOREndbd,GERDAndbd,Majoranandbd,EXO200ndbd,KamLANDZenndbd}, which, however, has never been observed~\cite{nohayndbd}, thus indicating that, even if neutrinos turn out to be associated to Majorana fields, the neutrinoless double beta decay is quite rare in nature. Alternative means to determine the Dirac or Majorana nature of neutrinos exist, such as the one given in Ref.~\cite{MRS}. Further differences among these approaches emerge at different levels. For instance, the Pontecorvo-Maki-Nakagawa-Sakata (PMNS) neutrino mixing matrix~\cite{MNSmatrix,Pontecorvomatrix}, here denoted by $U_\nu$, has either one complex phase, if neutrinos are Dirac like, or three complex phases, when Majorana neutrinos are assumed. The electromagnetic properties of the neutrinos crucially depend on whether they are Dirac or Majorana fermions~\cite{BGS}. While all diagonal electromagnetic moments of Dirac neutrinos are allowed, Majorana neutrinos do not have neither diagonal magnetic moment nor diagonal electric dipole moment, though their transition moments can be nonzero. It is well known that the sets of Feynman rules corresponding to each of such frameworks also differ of each other~\cite{DEHK,GlZr}. In general, the Majorana description comes along with more flexibility, thus usually allowing for a larger number of Feynman diagrams contributing to some given physical process, in comparison with the Dirac case. This turns out to be the case of the phenomenological calculation executed for the present investigation. A relevant observation, regarding the ${\cal L}_{\rm BSM}$ tree-level coupling of the $Z$ boson to neutrinos, is opportune: from Eq.~(\ref{LNC}), we express the neutral-currents Lagrangian for such couplings as ${\cal L}_{Z\nu\nu}=\sum_{k=1}^6\sum_{j=1}^6{\cal L}_{Zn_kn_j}$, with
\begin{equation}
{\cal L}_{Zn_kn_j}=-iZ_\mu\overline{n_k}\,\Gamma^\mu_{kj}n_j.
\end{equation}
Here, $n_k$ denotes a neutrino field, either light or heavy, where $n_1=\nu_1$, $n_2=\nu_2$, $n_3=\nu_3$, $n_4=N_1$, $n_5=N_2$, and $n_6=N_3$. On the other hand, $\Gamma^\mu_{kj}$ is a $4\times4$ matrix, defined in the space generated by the Dirac gammas, which, according to Eq.~(\ref{LNC}), depends on the $6\times6$ matrix ${\cal C}$, given in Ref.~\cite{Pilaftsis} and discussed in Appendix~\ref{App1}. If neutrinos were assumed to be Dirac fermions, the only Feynman rule for $Z\overline{n_k}n_j$ would be
\begin{equation}
\begin{gathered}
\vspace{-0.2cm}
\includegraphics[width=2.6cm]{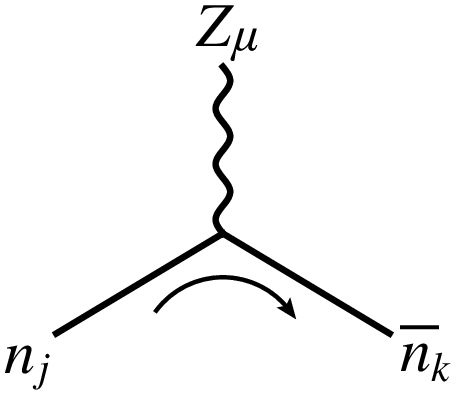}
\end{gathered}
=\Gamma^\mu_{kj}.
\label{Diracvertex}
\end{equation}
However, since the neutrinos considered for the present work are of Majorana type, an extra Feynman rule for $Z\overline{n_k}n_j$ is to be considered~\cite{DEHK}:
\begin{equation}
\begin{gathered}
\vspace{-0.2cm}
\includegraphics[width=2.6cm]{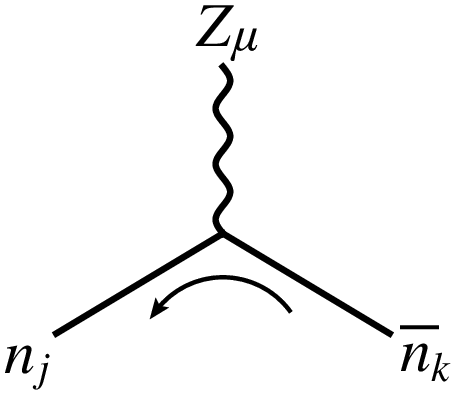}
\end{gathered}
=\Gamma'^\mu_{kj}=C\,\Gamma^{\mu{\rm T}}_{jk}C^{-1}.
\label{Majoranavertex}
\end{equation}
where $\Gamma^{\mu {\rm T}}_{jk}$ is the transpose matrix of $\Gamma^{\mu}_{jk}$ and $C$ is the aforementioned charge-conjugation matrix. While fermion number flow in Feynman diagrams featuring Majorana fermions does not apply, the authors of Ref.~\cite{DEHK} advise about the use of some fermion flow to set an orientation for fermion chains. The fermion flow in Eqs.~(\ref{Diracvertex}) and (\ref{Majoranavertex}) has been indicated by the arrows located off the neutrino lines, so establishing a distinction among these vertices.
\\

The $WWZ$ vertex-function contribution which we aimed to calculate comprises three sorts of contributing diagrams, which we gather into three terms. The vertex function, $\Gamma^{WWZ}_{\sigma\rho\mu}$, is then expressed as
\begin{equation}
\Gamma^{WWZ}_{\sigma\rho\mu}=\Gamma^{lln}_{\sigma\rho\mu}+\Gamma^{nnl}_{\sigma\rho\mu}+\Gamma^{nnh}_{\sigma\rho\mu},
\end{equation}
with the partial-amplitude contributions $\Gamma^{lln}_{\sigma\rho\mu}$, $\Gamma^{nnl}_{\sigma\rho\mu}$, and $\Gamma^{nnh}_{\sigma\rho\mu}$ given by
\begin{equation}
\Gamma^{lln}_{\sigma\rho\mu}=
\sum_{k=1}^6\sum_\alpha
\begin{gathered}
\vspace{-0.2cm}
\includegraphics[width=2cm]{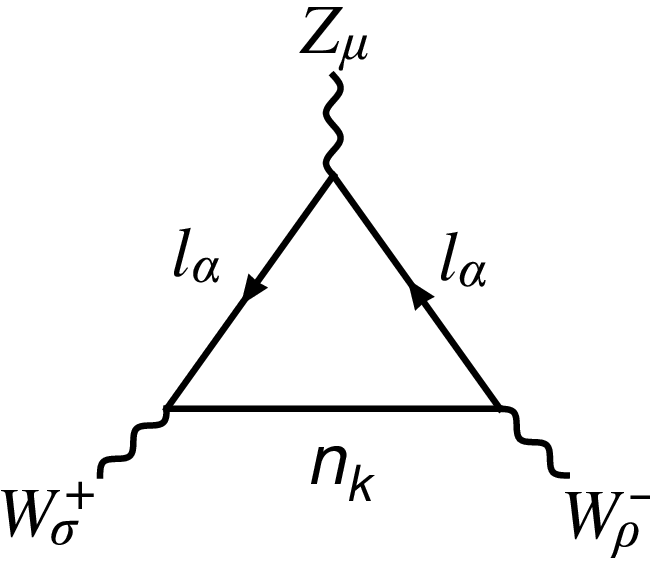}
\end{gathered}
,
\label{amp1}
\end{equation}
\begin{equation}
\Gamma^{nnl}_{\sigma\rho\mu}=
\sum_{k=1}^6\sum_{j=1}^6\sum_\alpha
\Bigg(
\begin{gathered}
\vspace{-0.2cm}
\includegraphics[width=2cm]{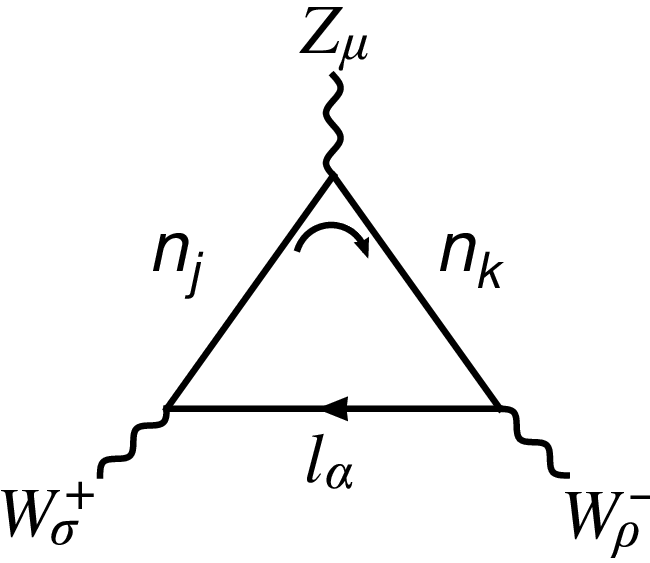}
\end{gathered}
+
\begin{gathered}
\vspace{-0.2cm}
\includegraphics[width=2cm]{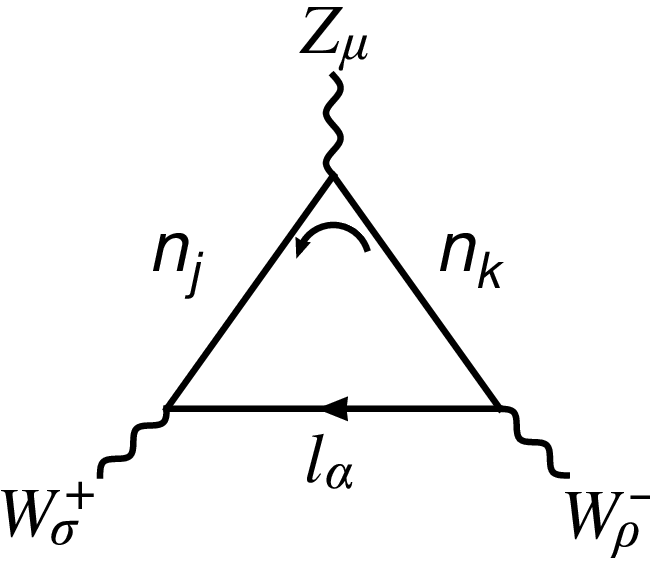}
\end{gathered}
\Bigg),
\label{amp2}
\end{equation}
\begin{equation}
\Gamma^{nnh}_{\sigma\rho\mu}=
\sum_{k=1}^6\sum_{j=1}^6
\Bigg(
\begin{gathered}
\vspace{-0.2cm}
\includegraphics[width=1.8cm]{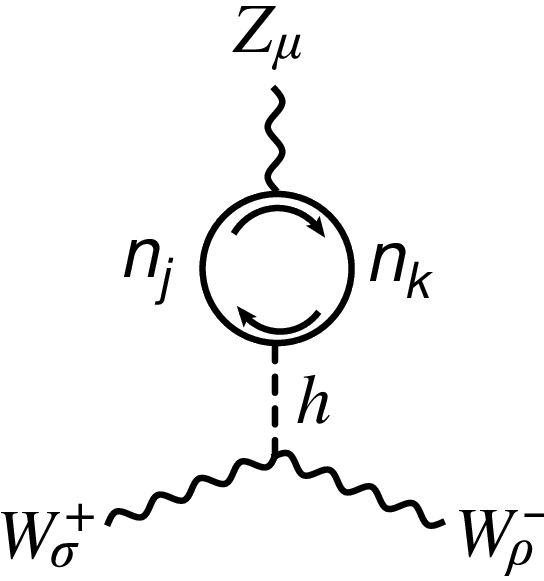}
\end{gathered}
+
\begin{gathered}
\vspace{-0.2cm}
\includegraphics[width=1.8cm]{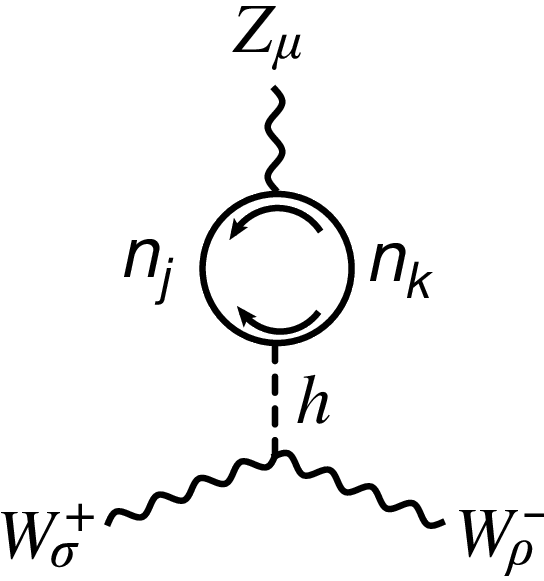}
\end{gathered}
\Bigg).
\label{amp3}
\end{equation}
Regarding Eq.~(\ref{amp2}), it shows two types of triangle diagrams, differing of each other solely by their vertices $Z\overline{n_k}n_j$, as the corresponding fermion flows point in opposite directions. The first of them is a Dirac-type diagram, that is, the only Feynman diagram which one would have to consider if the neutrinos were described by Dirac fields. The second diagram is a Majorana-type diagram, which does not occur in the Dirac case, but whose presence is allowed as long as Majorana fermions are assumed, and which has to be summed together with the Dirac-type diagram. It turns out that the properties of the charge-conjugation matrix $C$ and the Hermiticity of the $6\times6$ matrix ${\cal C}$, defined in Appendix~\ref{App1}, conspire to yield
\begin{equation}
\begin{gathered}
\vspace{-0.2cm}
\includegraphics[width=2cm]{WWZ2}
\end{gathered}
=
\begin{gathered}
\vspace{-0.2cm}
\includegraphics[width=2cm]{WWZ2M}
\end{gathered},
\end{equation}
so the analytical expressions of the Majorana-type diagrams and their alike Dirac-type diagrams turn out to coincide. Something similar happens with the contributing diagrams displayed in Eq.~(\ref{amp3}), where the first diagram is of Dirac type, whereas the second one is of Majorana type. These diagrams include a virtual Higgs-boson line, labeled by $h$. The Dirac-type diagram involves two fermion-flow arrows, with each of them corresponding to one of the two vertices, $Z\overline{n_k}n_j$ and $h\overline{n_j}n_k$, comprised by the fermion loop. The direction of any of these fermion flows continues the direction of the other. This is in contrast with the second diagram shown in this equation, as the fermion flows of the vertices $Z\overline{n_k}n_j$ and $h\overline{n_j}n_k$ point in opposite directions. This can be understood in two manners: (1) the vertex $Z\overline{n_k}n_j$ is of Majorana type, thus corresponding to what is shown in Eq.~(\ref{Majoranavertex}), while $h\overline{n_k}n_j$ is left as a Dirac-type vertex; or (2) the vertex $h\overline{n_j}n_k$ is the one to be inserted as a Majorana-type vertex, but $Z\overline{n_k}n_j$ is kept Dirac type. Just as it happened with the contributing diagrams in $\Gamma^{nnl}_{\sigma\rho\mu}$, we find that the relation
\begin{equation}
\begin{gathered}
\vspace{-0.2cm}
\includegraphics[width=2cm]{WWZ3}
\end{gathered}
=
\begin{gathered}
\vspace{-0.2cm}
\includegraphics[width=2cm]{WWZ3M},
\end{gathered}
\end{equation}
among Dirac-type and Majorana-type diagrams is fulfilled. Finally, we find it worth commenting that none of the diagrams in Eqs.~(\ref{amp2}) and (\ref{amp3}) have analogues in the context of the gauge vertex $WW\gamma$, as the electromagnetic field does not couple at tree level neither to neutrinos nor to the Higgs field. 
\\

Whenever a loop calculation is performed, ultraviolet divergences might come about, thus calling for a regularization method and a renormalization scheme. The superficial degree of divergence of the diagrams contributing to $\Gamma^{lln}_{\sigma\rho\mu}$ and $\Gamma^{nnl}_{\sigma\rho\mu}$, exhibited in Eqs.~(\ref{amp1}) and (\ref{amp2}), is 1, which means that the advent of linear divergences from these contributions can be anticipated. Moreover, the diagrams of Eq.~(\ref{amp3}), constituting the $\Gamma^{nnh}_{\sigma\rho\mu}$ contribution, are expected to produce ultraviolet divergences growing as large as quadratically, since their superficial degree of divergence is 2. We utilize the method of dimensional regularization~\cite{BoGi,tHooVe}, in which the dimension of spacetime is assumed to be $D\ne4$. Then, by analytic continuation, the complex quantity $\epsilon=4-D$ is defined, with $\epsilon\to0$. As part of the implementation of dimensional regularization, we carry out the change $\int \frac{d^4k}{(2\pi)^4}\to\mu_{\rm R}^{4-D}\int\frac{d^Dk}{(2\pi)^D}$ to all the 4-momentum integrals associated to loops in diagrams. Here, $\mu_{\rm R}$ is the renormalization scale, which has units of mass so that the factor $\mu_{\rm R}^{4-D}$ leaves the units of the loop integrals the same as they were in 4 dimensions. To handle the amplitudes given by the contributing diagrams, we follow the Passarino-Veltman tensor-reduction method~\cite{PaVe,DeSt}, which we carry out by usage of the software packages \textsc{FeynCalc}~\cite{SMO1,SMO2,MBD} and \textsc{Package-X}~\cite{Patel}, implemented in \textsc{Mathematica}, by Wolfram. After full implementation of the Passarino-Veltman method, all the AC form-factor contributions, displayed in Eqs.~(\ref{Gammaeven})-(\ref{Gammaodd}), are found to be functions on a variety of masses: masses of neutrinos $m_{n_j}$, both light and heavy; charged-lepton masses $m_\alpha$; and the mass of the $W$ boson, $m_W$. Provided the vertex $WWZ$ has been calculated with the external $Z$ boson taken off shell, the analytic expressions of the contributions also depend on the squared momentum of the corresponding external line. Recall that we have denoted $s=(2q)^2$, in accordance with the conventions set in Fig.~\ref{WWZ}. This mass and $s$ dependence also determines Passarino-Veltman scalar functions~\cite{tHooVescfunc} in terms of which the analytic expressions are given once the tensor-reduction method has been executed. In particular, these NP contributions depend on 1-point, 2-point, and 3-point scalar functions, defined as~\cite{PaVe,tHooVescfunc}
\begin{widetext}
\begin{equation}
A_0(m_0^2)=\frac{(2\pi)^{4-D}}{i\pi^2}\int d^Dk\frac{1}{k^2-m_0^2},
\end{equation}
\begin{equation}
B_0(p_1^2,m_0^2,m_1^2)=\frac{(2\pi)^{4-D}}{i\pi^2}\int d^Dk\frac{1}{\big( k^2-m_0^2 \big)\big( (k+p_1)^2-m_1^2 \big)},
\end{equation}
\begin{equation}
C_0(p_1^2,(p_1-p_2)^2,p_2^2,m_0^2,m_1^2,m_2^2)=\frac{(2\pi)^{4-D}}{i\pi^2}\int d^Dk\frac{1}{\big( k^2-m_0^2 \big)\big( (k+p_1)^2-m_1^2 \big)\big( (k+p_2)^2-m_2^2 \big)}.
\label{PaVeC0}
\end{equation}
\end{widetext}
Aiming at the general structure of the $WWZ$ vertex function, given in Eqs.~(\ref{Gammaeven}) and (\ref{Gammaodd}), we split the NP contribution $\Gamma^{WWZ}_{\sigma\rho\mu}$ into $CP$-even and $CP$-odd terms. The contributions to the $CP$-invariant AC $\Delta Q$ and $\Delta\kappa$ are found to be ultraviolet finite, and the same goes for $f_1$. The only ultraviolet divergence which remains lies in $g_1$, which is expected, as this coupling has a tree-level counterpart, so it can be renormalized. With respect to the $CP$-violating contributions, care has to be taken when calculating them, since the implementation of dimensional regularization in the presence of the chirality matrix, $\gamma_5$, may yield spurious contributions~\cite{Jegerlehner}. For instance, in the case of amplitudes involving external photon lines, apparent violations of Ward identities can be generated, thus misleadingly indicating the presence of anomalies~\cite{Adler,BeJa,Fujikawa} in anomaly-free theoretical frameworks. Of course, note that an issue so delicate does not occur in the present calculation, in which fulfillment of Ward identities is not a requirement. In practice, the main issue lies in Dirac-matrix traces ${\rm tr}\{ \gamma^\mu\gamma^\nu\gamma^\alpha\gamma^\beta\gamma_5 \}$, of four gamma matrices and the chirality matrix. Such traces are inconsistently set to 0 in $D$ dimensions in the so-called ``naive dimensional regularization'' scheme, where $\{ \gamma^\mu,\gamma_5 \}=0$, for all $\mu=0,1,2,3,4,\ldots,D-1$, is assumed. Another approach is the 't Hooft-Veltman one~\cite{tHooVe}, in which $\{ \gamma^\mu,\gamma_5 \}=0$ for $\mu=0,1,2,3$, and $[ \gamma,\gamma^\mu ]=0$ for $\mu=4,5,\ldots,D-1$. This allows for ${\rm tr}\{ \gamma^\mu\gamma^\nu\gamma^\alpha\gamma^\beta\gamma_5 \}$ to be nonzero, though fake-anomaly terms might be generated in this way. Variants of the 't Hooft-Veltman procedure are available~\cite{BrMa,CFH,AoTo,Bonneau}. To tackle the issue of the Dirac-matrix traces in $D$ dimensions, we have left the traces ${\rm tr}\{ \gamma^\mu\gamma^\nu\gamma^\alpha\gamma^\beta\gamma_5 \}$ unsolved during the computation in $D$ spacetime dimensions. After processing the analytical contributions, we have verified that all the coefficients of such Dirac traces are finite in the ultraviolet sense, and, after that, we have evaluated such traces in $4$ spacetime dimensions. It is worth commenting that the algebraic procedure leaded to an exact cancellation of the $CP$-odd contribution $\Delta\tilde{Q}$, which happens as we describe next. Among the whole set of terms involving traces ${\rm tr}\{ \gamma^\mu\gamma^\nu\gamma^\alpha\gamma^\beta\gamma_5 \}$, we notice that the $CP$-odd anomaly $\Delta\tilde{Q}$ can only get contributions from the combination $ig_Z\big(w_1\, \epsilon_{\mu\rho\alpha\beta}p^\alpha q^\beta q_\sigma+w_2\,\epsilon_{\rho\sigma\alpha\beta}p^\alpha q^\beta p_\mu\big)$, with $w_1$ and $w_2$ given in terms of masses and $s$. With the aid of Schouten identities~\cite{NeVe}, we write this sum as
\begin{eqnarray}
&&
w_1\, \epsilon_{\mu\rho\alpha\beta}p^\alpha q^\beta q_\sigma+w_2\,\epsilon_{\rho\sigma\alpha\beta}p^\alpha q^\beta p_\mu
\nonumber \\ &&
\hspace{2.5cm}
=\frac{4\Delta\tilde{Q}}{m_W^2}q_\rho\epsilon_{\sigma\mu\alpha\beta}p^\alpha q^\beta+\cdots,
\end{eqnarray}
where the contribution $\Delta\tilde{Q}=\frac{m_W^2(2w_2-w_1)}{4}$ has been identified. The relation $w_1=2w_2$ tuns out to hold, thus implying the vanishing of the $\Delta\tilde{Q}$ AC. A similar cancellation of a NP contribution from Majorana neutrinos to $\Delta\tilde Q$, at the one-loop level, was reported in Ref.~\cite{BuPi}. Regarding the remaining nonzero $CP$-odd terms of the $WWZ$ vertex, that is, $\Delta\tilde{\kappa}$, $\tilde{f}_1$, and $\tilde{f}_2$ in Eq.~(\ref{Gammaodd}), we arrived at the conclusion that all of them are free of ultraviolet divergences. 
\\

From now on, our discussion is entirely focused in the NP contributions $\Delta\kappa$, $\Delta Q$, and $\Delta\tilde{\kappa}$. Let us comment that the vertex-function term $\Gamma^{nnh}_{\sigma\rho\mu}$, Eq.~(\ref{amp3}), does not generate contributions to these AC. In contraposition, the terms $\Gamma^{lln}_{\sigma\rho\mu}$ and $\Gamma^{nnl}_{\sigma\rho\mu}$, given by Eqs.~(\ref{amp1}) and (\ref{amp2}), are the ones which completely define the NP contributions of interest. They can be generically written as
\begin{eqnarray}
&&
\Delta\zeta=
\sum_{k=1}^6\sum_\alpha
|{\cal B}_{\alpha k}|^2\Delta\zeta^{(1)}_{\alpha k}
\nonumber \\ &&
\hspace{0.1cm}
+\sum_{j=1}^6\sum_{k=1}^6\sum_\alpha{\cal B}_{\alpha k}{\cal B}^*_{\alpha j}\big( {\cal C}_{kj}\Delta\zeta^{(2)}_{\alpha kj}+{\cal C}^*_{kj}\Delta\zeta^{(3)}_{\alpha kj}\big),
\label{genanomaly}
\end{eqnarray}
where $\zeta=\kappa, Q, \tilde{\kappa}$. Moreover, $\Delta\zeta^{(1)}_{\alpha k}$, $\Delta\zeta^{(2)}_{\alpha kj}$, and $\Delta\zeta^{(3)}_{\alpha kj}$ are functions of masses $m_W$, $m_{n_j}$, $m_\alpha$, and on the squared external $Z$-boson momentum $s$. The first term in this equation, which involves the factors $\Delta\zeta^{(1)}_{\alpha k}$, comes from $\Gamma^{lln}_{\sigma\rho\mu}$, whereas the second term, in which $\Delta\zeta^{(2)}_{\alpha kj}$ and $\Delta\zeta^{(3)}_{\alpha kj}$ appear, is generated by $\Gamma^{nnl}_{\sigma\rho\mu}$. Let us remark that $\Delta Q^{(3)}_{\alpha kj}=0$, so $\Delta Q$, by contrast with the other AC contributions, lacks terms proportional to ${\cal C}^*_{kj}$. We find that the neutrino-indices symmetry property $\Delta\zeta^{(a)}_{\alpha kj}=\Delta\zeta^{(a)}_{\alpha jk}$, where $a=2,3$, is fulfilled in the case of the $CP$-even contributions. Notice that the $\Delta\zeta^{(a)}$ factors are complex-valued quantities. To grasp this assertion, take into account that the diagrams contributing to $\Gamma^{lln}_{\sigma\rho\mu}$, Eq.~(\ref{amp1}), include a vertex which connects the external $Z$-boson line to a couple of charged-lepton lines. Since the relation $4m^2_\alpha<s$, among the squared momentum of the virtual $Z$ boson line and the mass of any SM charged lepton, must hold for the $W$-boson pair production to be allowed, the corresponding amplitude is complex valued. Moreover, as far as the contributing diagrams comprising $\Gamma_{\sigma\rho\mu}^{nnl}$, Eq.~(\ref{amp2}), are concerned, the same line of reasoning applies when the virtual-neutrino lines, which are connected to the external $Z$-boson line, correspond to light neutrinos.
\\


\section{Estimation of contributions to AC and discussion}
\label{NumPheno}
Now we perform an estimation of the contributions whose analytic calculation we discussed throughout the previous section. The estimation and analysis of $CP$-even and $CP$-odd AC are addressed in separate subsections.
\\

In Appendix~\ref{App1}, the matrices ${\cal B}$ and ${\cal C}$, on whose entries the NP contributions $\Delta\kappa$, $\Delta Q$, and $\Delta\tilde\kappa$ depend, are defined and discussed. Notice that these matrices introduce several parameters. With the sole purpose of achieving an estimation of contributions, we take a pragmatic approach by writing the matrix $\xi$, used in Eq.~(\ref{Uapprox}) to parametrize the unitary diagonalization matrix ${\cal U}$, as
\begin{equation}
\xi=\hat{\rho}X,
\label{rhoXdef}
\end{equation}
where $\hat\rho$ is a number, both real and positive. We think of $\hat\rho$ as being equal to the modulus of the largest entry of $\xi$. Therefore $X$ is a $3\times3$ sized complex matrix with the largest modulus among all its entries being 1. The investigations developed in Refs.~\cite{MMNS} and \cite{NoSa}, centered on contributions from Majorana neutrinos to $WW\gamma$ and $ZZZ$, have profited from this procedure to get estimations and then analyze the contributions from the neutrino-mass model of Ref.~\cite{Pilaftsis}. Implementation of Eq.~(\ref{rhoXdef}) in Eqs.~(\ref{Binxi}) and (\ref{Cinxi}) yields the expressions
\begin{equation}
{\cal B}=
\left(
\begin{array}{cc}
V^\ell\big( {\bf 1}_3+\hat{\rho}^2XX^\dag \big)^{-\frac{1}{2}}
&
V^\ell\hat\rho X\big( {\bf 1}_3+\hat{\rho}^2X^\dag X \big)^{-\frac{1}{2}}
\end{array}
\right),
\end{equation}
\begin{equation}
{\cal C}=
\left(
\begin{array}{cc}
\big({\bf 1}_3+\hat{\rho}^2XX^\dag\big)^{-1} & \hat\rho\big( {\bf 1}_3+\hat{\rho}^2XX^\dag \big)^{-1}X
\vspace{0.2cm}
\\
\hat\rho X^\dag\big( {\bf 1}_3+\hat{\rho}^2XX^\dag \big)^{-1} & \hat\rho^2X^\dag\big( {\bf 1}_3+\hat\rho^2XX^\dag \big)^{-1}X
\end{array}
\right).
\end{equation}
Then, our estimations are carried out by considering the matrix texture 
\begin{equation}
X=e^{i\varphi}\cdot{\bf 1}_3.
\label{Xtexture1}
\end{equation}
Moreover, we take $V^\ell=U_\nu$. The PMNS matrix $U_\nu$ is written, in the case of Majorana neutrinos, as $U_\nu=U_{\rm D}\,U_{\rm M}$, with $U_{\rm M}={\rm diag}\big( 1,e^{i\phi_1},e^{\i\phi_2} \big)$, where $\phi_1$ and $\phi_2$ are the so-called Majorana phases, which are exclusive of the Majorana-neutrinos framework. Let us comment that our estimations are independent of the Majorana phases. The matrix factor $U_{\rm D}$, on the other hand, is conventionally parametrized as
\begin{widetext}
\begin{equation}
U_{\rm D}=
\left(
\begin{array}{ccc}
c_{12}c_{13} & s_{12}s_{13} & s_{13} e^{-i\delta_{\rm D}}
\vspace{0.2cm}
\\
-s_{12}c_{23}-c_{12}s_{23}s_{13}e^{i\delta_{\rm D}} & c_{12}c_{23}-s_{12}s_{23}s_{13}e^{i\delta_{\rm D}} & s_{23}c_{13}
\vspace{0.2cm}
\\
s_{12}s_{23}-c_{12}c_{23}s_{13}e^{i\delta_{\rm D}} & -c_{12}s_{23}-s_{12}c_{23}s_{13}e^{i\delta_{\rm D}} & c_{23}c_{13}
\end{array}
\right).
\end{equation}
\end{widetext}
This matrix, which depends on the mixing angles $\theta_{12}$, $\theta_{23}$, and $\theta_{13}$, is given in terms of sines $s_{jk}=\sin{\theta_{jk}}$ and cosines $c_{jk}=\cos{\theta_{jk}}$. It also includes the parameter $\delta_{\rm D}$, known as the ``Dirac phase''. For our numerical estimations, we consider the following values for the $U_{\rm D}$ mixing angles: $s^2_{12}=0.307\pm0.013$, $s^2_{23}=0.546\pm0.0021$, and $s^2_{13}=0.0220\pm0.0007$. These are the best values provided by the Particle Data Group~\cite{PDG} (PDG), based on measurements from diverse experimental collaborations. For $\theta_{12}$, a measurement by the Super-Kamiokande Collaboration, reported in Ref.~\cite{SuperKamiokandemixing} was taken by the PDG. Furthermore, the PDG $\theta_{23}$ value comes from measurements by T2K~\cite{T2K2}, Minos+~\cite{MinosPlusmixing}, NOvA~\cite{Novamixing}, IceCube~\cite{IceCubemixing}, and Super-Kamiokande~\cite{SuperKamiokandeanothermixing}. Concerning the $\theta_{13}$ mixing angle, the PDG considered measurements reported by Double Chooz~\cite{DoubleChoozmixing}, RENO~\cite{Renomixing1,Renomixing2}, and Daya Bay~\cite{DayaBaymixing1,DayaBaymixing2}. For the Dirac phase we use $\delta_{\rm D}=-\frac{\pi}{2}$, which has been favored by the T2K Collaboration~\cite{T2K1,T2K2}.
\\

While the measurement of neutrino oscillations has led to the conclusion that light neutrinos are massive, as opposed to what is assumed in the SM, this phenomenon does not provide information on the absolute neutrino-mass scale. Cosmological observations have yielded a stringent limit on the sum of light-neutrino masses, namely, $\sum_jm_{\nu_j}<0.12\,{\rm eV}$, at $95\%\,{\rm C.L.}$~\cite{SDSSexperiment,Planckexperiment}. Recently, the KATRIN Collaboration announced the achievement of an upper limit on the effective electron anti-neutrino mass, defined by $m_{\nu_e}^{2({\rm eff})}=\sum_j|(U_\nu)_{ej}|^2m_{\nu_j}^2$~\cite{KATRIN}, from which the upper bound $0.8\,{\rm eV}$, at $90\%\,{\rm C.L.}$, on neutrino mass was established. Remarkably, this limit does not rely on any cosmological assumption and it applies independently of whether neutrinos are Majorana or Dirac fermions. By virtue of the smallness of light-neutrino masses, and for the sake of practicality, for our estimations we take them all to be the same, so $m_{\nu_j}=m_\nu$ for all $j=1,2,3$. The neutrino-mass model under consideration for the present investigation, given in Ref.~\cite{Pilaftsis}, established that a quasi-degenerate heavy-neutrino mass spectrum is imperious in order to get light-neutrino masses consistent with current upper bounds. With this in mind, we set all the heavy-neutrino masses to be equal to each other\footnote{While the near-degenerate heavy-neutrino spectrum is imposed by the neutrino model~\cite{Pilaftsis}, we have verified that assuming mass non-degeneracy does not produce sizable deviations from our estimations.}, that is, $m_{N_j}=m_N$, for all $j=1,2,3$. With respect to the heavy-neutrino mass $m_N$ we use for our estimations, we refer the reader to Ref.~\cite{CMShnl}, where a study, performed by the CMS Collaboration, on the mass of heavy neutral leptons and its relation with their mixing to light neutrinos is discussed. That work presents a couple of remarkable graphs which show, independently of models, exclusion regions in the planes $\big( |{\cal B}_{eN}|^2,m_N \big)$ and $\big( |{\cal B}_{\mu N}|^2,m_N \big)$, in accordance with LHC data. We have profited from such graphs, using them to select the values $\hat{\rho}=0.58,0.65$, and also to determine that these choices are consistent with the constraint $700\,{\rm GeV}\leqslant m_N$, on heavy-neutrino mass. We take this minimum heavy-neutrino mass value as a reference for our forthcoming estimations. A more detailed discussion on these choices for the $\hat\rho$ parameter is presented below. As we show later, the $X$-matrix texture, given in Eq.~(\ref{Xtexture1}), and the $\hat\rho$ parameter, taken to abide by the CMS constraints of Ref.~\cite{CMShnl}, provide an estimation of the $WWZ$ contribution, consistent with experimental data. Bear in mind, however, that other possibilities exist. The Hermitian matrix $\eta=\frac{1}{2}\xi\xi^\dag$ has been customarily used to characterize non-unitary effects in light neutrino mixing~\cite{FGLY}. Since ${\cal B}_\nu=U_\nu\big( {\bf 1}_3-\eta \big)$ and ${\cal B}_N=\hat\rho\,U_\nu\big( {\bf 1}_3-\eta \big)X$, non-unitary effects are, according to Eq.~(\ref{genanomaly}), involved in the NP contributions to $WWZ$ discussed through the present paper. Moreover, since $\eta=\frac{1}{2}\hat\rho^2 XX^\dag$, constraints on $\eta$ can be used to impose restrictions on $\hat\rho$. For instance, detailed analyses on non-unitary effects were carried out in Refs.~\cite{FHL,BFHLMN}, where global fits leaded to stringent constraints on the matrix $\eta$, with upper bounds on the moduli of the entries of this matrix ranging within $\sim10^{-7}$ to $10^{-3}$. These limits would set severe restrictions on the $\hat\rho$ parameter, therefore yielding an extra suppression on the AC contributions, which we estimate to amount to about two orders of magnitude. Nonetheless, keep in mind that the implementation of these bounds would require further assumptions on the structure of the matrices $m_{\rm D}$ and $m_{\rm M}$ (see Appendix~\ref{App1}), associated to the presence of an underlying lepton-number symmetry, only slightly violated.
\\

The $WWZ$ vertex can be probed by hadron colliders. Based on the Tevatron accelerator, the D0 detector was utilized to set an upper bound of order $10^{-1}$ on $\Delta\kappa$ and also an upper limit on $\Delta Q$, of order $10^{-2}$, with both constraints given at $95\%\,{\rm C.L.}$, which was reported in Ref.~\cite{D0onWWZ}, were the D0 Collaboration considered $WW$, $WZ$, and $W\gamma$ production from proton-antiproton collisions taking place at a CME of $1.96\,{\rm TeV}$. The ATLAS and CMS Collaborations, at the Large Hadron Collider, have also explored this gauge vertex. In Refs.~\cite{ATLASonWWZ, CMSonWWZ}, these collaborations reported upper limits of order $10^{-2}$, at $95\%\,{\rm C.L.}$, on both $\Delta\kappa$ and $\Delta Q$, by analyzing $WW$ and $WZ$ production from proton-proton collisions at a CME of $8\,{\rm TeV}$. An ulterior analysis of $W\gamma$ production from proton-proton collisions at a CME of $13\,{\rm TeV}$ was recently performed by the CMS Collaboration, which established a constraint of order $10^{-3}$ on the $\Delta Q$ analogue that corresponds to the vertex $WW\gamma$~\cite{CMSonWWZimproved}. Experimental works on the AC of the $WWZ$ and $WW\gamma$ gauge vertices often assume that $\Delta Q$, of the $WWZ$ vertex, coincides with such an analogue, thus implying a direct translation of the aforementioned upper limit into an improved upper bound on the $WWZ$-vertex form factor $\Delta Q$. Keep in mind, however, that, as pointed out in Ref.~\cite{EllWu}, such an assumption makes sense as a good approximation as long as the scale of NP lies beyond $3\,{\rm TeV}$. This gauge vertex can also be studied by means of electron-positron colliders, which offer a cleaner environment, in comparison with hadron colliders. Whenever the energy threshold $2m_W$ is surpassed by the CME of some electron-positron collider, such a machine shall be able to produce $W$-boson pairs from the process $e^+e^-\to W^+W^-$, which receives contributions, through $s$-channel diagrams mediated by a virtual $Z$ boson, from the $WWZ$ vertex. The Large Electron-Positron (LEP) collider, which terminated operations since 2000 for its installations to become part of the Large Hadron Collider, used to be a circular electron-positron accelerator endowed with four detectors: ALEPH, DELPHI, L3, and OPAL. The LEP accelerator started colliding $e^+e^-$ pairs at a CME of $\sim91\,{\rm GeV}$, then being upgraded to eventually reach the threshold for $W$-boson pair production, finally reaching its maximum energy at $209\,{\rm GeV}$. In 2013, the collaborations which worked at the four LEP detectors put together data of $WW$ production from $e^+e^-$ collisions at CMEs ranging from $130\,{\rm GeV}$ to $209\,{\rm GeV}$, and then performed a high-precision analysis of AC, from which bounds of order $10^{-2}$ on both $\Delta\kappa$ and $\Delta Q$ were established~\cite{LEPsuperbound}. The DELPHI Collaboration used data taken at different $e^+e^-$ CMEs, ranging from $189\,{\rm GeV}$ to $209\,{\rm GeV}$, to set constraints on the $CP$-odd AC $\Delta\tilde\kappa$ and $\Delta\tilde Q$ of order $10^{-2}$, at best~\cite{DelphiCPodd}. Future electron-positron colliders are in plans, among which the ILC has called for attention for years. The ILC Technical Design Report~\cite{ILC2} has presented estimations and analyses on the expected sensitivity of this devise, then arriving at the conclusion that the ILC will be able to set upper bounds on $\Delta\kappa$ and $\Delta Q$ of order $10^{-4}$ from $W$-pair production at CMEs of $\sqrt{s}=500\,{\rm GeV}$ and $\sqrt{s}=800\,{\rm GeV}$. That reference also estimated upper bounds on the $CP$-violating factors $\Delta\tilde{\kappa}$ and $\Delta\tilde{Q}$, from collisions at CMEs of $500\,{\rm GeV}$ and $800\,{\rm GeV}$, asseverating that the former quantity would be bounded up to order $10^{-2}$, whereas the latter is expected to be constrained up to order $10^{-4}$. Since we utilize the upper bounds of Ref.~\cite{ILC2} for our estimations, we display the absolute values of their numbers in Table~\ref{boundsbyILC}.
\begin{table}[ht]
\begin{tabular}{c|c|c}
\hline
Coupling & $\sqrt{s}=500\,{\rm GeV}$ & $\sqrt{s}=800\,{\rm GeV}$
\\ \hline
$|\Delta\kappa_{\rm ILC}|$ & $3.20\times10^{-4}$ & $1.90\times10^{-4}$
\\
$|\Delta Q_{\rm ILC}|$ & $1.34\times10^{-3}$ & $6.00\times10^{-4}$
\\
$|\Delta\tilde\kappa_{\rm ILC}|$ & $5.33\times10^{-2}$ & $5.77\times10^{-2}$
\\
$|\Delta\tilde Q_{\rm ILC}|$ & $1.5\times10^{-3}$ & $6.00\times10^{-4}$
\\ \hline
\end{tabular}
\caption{\label{boundsbyILC} Absolute values of expected ILC upper bounds on the $CP$-conserving AC $\Delta\kappa$ and $\Delta Q$, and the $CP$-violating AC $\Delta\tilde\kappa$ and $\Delta\tilde Q$, as estimated in Ref.~\cite{ILC2}.}
\end{table}
Another in-plans lepton accelerator is the  Circular Electron-Positron Collider, which aims at reaching bounds of order $10^{-4}$ on both $\Delta\kappa$ and $\Delta Q$~\cite{CEPConWWZ}.
\\

\subsection{CP even contributions}
\label{CPevensection}
Now we estimate and discuss the contributions from Majorana neutrinos, within the neutrino model of Ref.~\cite{Pilaftsis}, to the anomalies $\Delta\kappa$ and $\Delta Q$, both of them associated to $CP$-preserving NP. The one-loop contributions from the SM to $WWZ$ were addressed long ago in Ref.~\cite{AKLPS}, where a calculation of this vertex, together with $WW\gamma$, was executed, resulting in AC given as functions on the CME, with the masses of the Higgs boson and the top quark, not yet measured at the time, left as further parameters. Shortly after, this calculation was revisited in Ref.~\cite{PaPhi}, where the authors pointed out that the SM contributions bear gauge dependence, thus calling for a more careful treatment. A remarkable calculation by the authors of this reference, who followed the pinch technique~\cite{Cornwall,CoPa,Papavassiliou}, yielded results both well behaved and gauge independent. Contributions from the SM to these anomalies are $\Delta\kappa_{\rm SM}\sim10^{-3}$ and $\Delta Q_{\rm SM}\sim10^{-4}$~\cite{AKLPS,PaPhi}. It is worth commenting that $CP$-odd contributions $\Delta\tilde\kappa_{\rm SM}$ and $\Delta\tilde Q_{\rm SM}$, from the SM, have been found to be absent at one loop. Of course, loop effects, and thus AC, can also be generated by NP beyond the SM. For instance, in Ref.~\cite{AKM} the contributions from the MSSM were considered. In Ref.~\cite{MTTR}, a calculation of contributions from bileptons, emerged in the NP framework of the 331 model~\cite{PiPl,Frampton}, to $WWZ$ were calculated and estimated. The conclusion of the authors was that contributions as large as those from the SM are attainable as long as these bileptons are relatively light. A follow up of that work, Ref.~\cite{RTT}, considered a variant of the 331 model, which features right-handed neutrinos, and performed the calculation of the $WWZ$ vertex. Models of universal extra dimensions~\cite{ACD} have also been utilized to study $WWZ$. Refs.~\cite{FMNRT,LMMNTT} reported calculations of contributions to $\Delta\kappa$ and $\Delta Q$ from the 5-dimensional Yang-Mills theory and from the whole SM defined in 5 spacetime dimensions, achieving contributions around 2 orders of magnitude below those from the SM. The Georgi-Machacek model~\cite{GeMa} was considered in Ref.~\cite{UHT} to calculate $WWZ$ at one loop, from which estimations of $\Delta\kappa$ and $\Delta Q$ were achieved. 
\\

As we mentioned above, the $\Delta\kappa$ and $\Delta Q$ anomaly contributions are complex valued, so we execute our estimations by rather considering the moduli $|\Delta\kappa|$ and $|\Delta Q|$. With respect to the coefficients $\Delta\zeta^{(a)}$, which comprise Eq.~(\ref{genanomaly}), note that their neutrino indices run over the six neutrinos. Taking this into account, we can split these factors into
\begin{equation}
\Delta\zeta^{(1)}_{\alpha k}=
\left\{
\begin{array}{l}
\Delta\zeta^{(1)}_{\alpha\nu_j},\textrm{ if }k=1,2,3,
\vspace{0.2cm}
\\
\Delta\zeta^{(1)}_{\alpha N_j},\textrm{ if }k=4,5,6,
\end{array}
\right.
\end{equation}
\begin{equation}
\Delta\zeta^{(a)}_{\alpha ki}=
\left\{
\begin{array}{l}
\Delta\zeta^{(a)}_{\alpha\nu_j\nu_l},\textrm{ if }k=1,2,3,\textrm{ and }i=1,2,3,
\vspace{0.2cm}
\\
\Delta\zeta^{(a)}_{\alpha\nu_jN_l},\textrm{ if }k=1,2,3,\textrm{ and }i=4,5,6,
\vspace{0.2cm}
\\
\Delta\zeta^{(a)}_{\alpha N_j\nu_l},\textrm{ if }k=4,5,6,\textrm{ and }i=1,2,3,
\vspace{0.2cm}
\\
\Delta\zeta^{(a)}_{\alpha N_jN_l},\textrm{ if }k=4,5,6,\textrm{ and }i=4,5,6,
\end{array}
\right.
\label{Deltazeta23}
\end{equation}
where $\nu_j,\nu_l=\nu_1,\nu_2,\nu_3$ and $N_j,N_l=N_1,N_2,N_3$. Moreover, $a=2,3$ in Eq.~(\ref{Deltazeta23}). Now we wish to point out that, for any fixed $a=1,2,3$ and $\zeta=\kappa, Q$, these $\Delta\zeta^{(a)}$ factors differ of each other only by their dependence on both the charged-lepton masses $m_\alpha$ and the neutrino masses $m_{\nu_{j}}$ and $m_{N_{j}}$. For instance, the coefficients $\Delta\kappa^{(3)}_{e\,\nu_3N_1}$ and $\Delta\kappa^{(3)}_{\tau N_2N_3}$ share the same structure in the sense that the implementation of the changes $m_\tau\to m_e$, $m_{N_2}\to m_{\nu_3}$, and $m_{N_3}\to m_{N_1}$ on the former yields the latter, that is, $\Delta\kappa^{(3)}_{\tau N_2N_3}\to\Delta\kappa^{(3)}_{e\, \nu_3N_1}$. Since, as we explained above, we are taking $m_{\nu_j}=m_\nu$ and $m_{N_j}=m_N$, for all $j=1,2,3$, the complete sets of 78 different $\Delta\kappa^{(a)}$ factors and 42 coefficients $\Delta Q^{(a)}$, for any fixed lepton flavor $\alpha$, reduce to a set of 10 quantities in the case $\zeta=\kappa$ and a set of only 6 for $\zeta=Q$: $\Delta\zeta^{(1)}_{\alpha\nu}$, $\Delta\zeta^{(1)}_{\alpha N}$, $\Delta\zeta^{(a)}_{\alpha\nu\nu}$, $\Delta\zeta^{(a)}_{\alpha\nu N}$, $\Delta\zeta^{(a)}_{\alpha N\nu}$, and $\Delta\zeta^{(a)}_{\alpha NN}$, with $a=2,3$. Recall that $\Delta\zeta^{(2)}_{\alpha kj}$ and $\Delta\zeta^{(3)}_{\alpha kj}$ are symmetric with respect to their neutrino indices, thus meaning that $\Delta\zeta^{(a)}_{\alpha\nu N}=\Delta\zeta^{(a)}_{\alpha N\nu}$ holds. Using Eqs.~(\ref{rhoXdef})-(\ref{Xtexture1}), together with the consideration of quasi-degenerate neutrino-mass spectra for light neutrinos and heavy neutrinos, we express the $CP$-preserving anomalies as
\begin{eqnarray}
&&
\Delta\kappa=\sum_\alpha
\Big(\frac{1}{1+\hat\rho^2}
\big( \Delta\kappa^{(1)}_{\alpha\nu}+\hat{\rho}^2\Delta\kappa^{(1)}_{\alpha N} \big)
\nonumber \\ &&
\hspace{0.8cm}
+\frac{1}{(1+\hat\rho^2)^2}\big( \Delta\kappa^{(2)}_{\alpha\nu\nu}+\Delta\kappa^{(3)}_{\alpha\nu\nu} \big)
\nonumber \\ &&
\hspace{0.8cm}
+\frac{\hat\rho^4}{(1+\hat\rho^2)^2}\big( \Delta\kappa^{(2)}_{\alpha NN}+\Delta\kappa^{(3)}_{\alpha NN} \big)
\nonumber \\ &&
\hspace{0.8cm}
+\frac{2\hat\rho^2}{(1+\hat\rho^2)^2}\big( \Delta\kappa^{(2)}_{\alpha\nu N}+\cos( 2\varphi )\Delta\kappa^{(3)}_{\alpha\nu N} \big)
\Big),
\label{Dkdegenerate}
\end{eqnarray}
\begin{eqnarray}
&&
\Delta Q=\sum_{\alpha}
\Big(
\frac{1}{1+\hat\rho^2}
\big( \Delta Q^{(1)}_{\alpha\nu}+\hat{\rho}^2\Delta Q^{(1)}_{\alpha N} \big)
\nonumber \\ &&
\hspace{0.5cm}
+\frac{1}{(1+\hat\rho^2)^2}\Delta Q^{(2)}_{\alpha\nu\nu}+\frac{\hat\rho^4}{(1+\hat\rho^2)^2}\Delta Q^{(2)}_{\alpha NN}
\nonumber \\ &&
\hspace{0.5cm}
+\frac{2\hat{\rho}^2}{(1+\hat\rho^2)^2}\Delta Q^{(2)}_{\alpha\nu N}
\Big).
\label{DQdegenerate}
\end{eqnarray} 
Explicit expressions for the coefficients $\Delta\zeta^{(a)}$ are provided in Appendix~\ref{App3}. Among these two NP contributions, the only remaining dependence on the complex phase $\varphi$, introduced by the matrix texture displayed in Eq.~(\ref{Xtexture1}), resides in $\Delta\kappa$. The presence of this phase results from the interplay of the ${\cal B}$- and ${\cal C}$-matrix factors in ${\cal B}_{\alpha \nu}{\cal B}^*_{\alpha N}{\cal C}^*_{\nu N}\varpropto e^{-2i\varphi}$, which accompanies each $\Delta\kappa^{(3)}_{\alpha\nu N}$ contribution. This is in contrast with the factors ${\cal B}_{\alpha \nu}{\cal B}^*_{\alpha N}{\cal C}_{\nu N}$, associated to $\Delta\kappa^{(2)}_{\alpha\nu N}$ and in which the $\varphi$ dependence vanishes. Since $\Delta Q$ lacks $\Delta Q^{(3)}_{\alpha\nu N}$ contributing factors, any possible dependence on $\varphi$ is absent. We have found that the contributions from $\Delta\kappa^{(3)}_{\alpha\nu N}$ are indeed much smaller than those from the dominant terms, by about 10 order of magnitude, which means that the variation introduced by the $\varphi$-dependence is not relevant in practice.
\\

\begin{figure}[ht]
\center
\includegraphics[width=8.5cm]{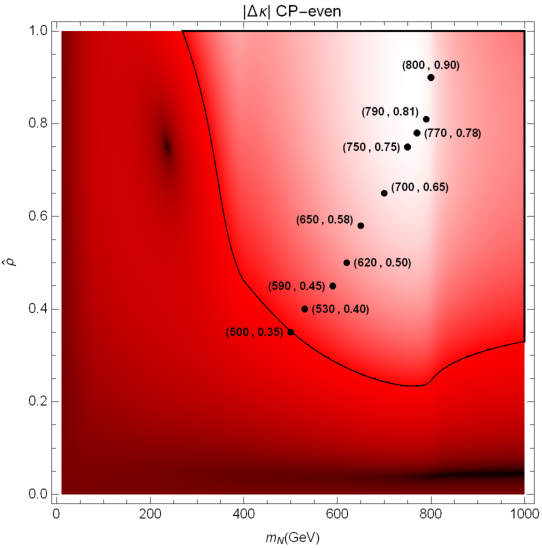}
\vspace{0.2cm} \\
\includegraphics[width=8.5cm]{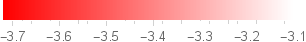}
\caption{\label{rhomasspars} One-loop AC contribution $|\Delta\kappa|$ from Majorana neutrinos in the $\big( m_N,\hat\rho \big)$ plane, plotted in base-10 logarithmic scale. The solid curve represents the boundary for the region situated in the upper-right sector of the graph, where the NP $|\Delta\kappa|$ contribution could be sensed by ILC~\cite{ILC2}. Points $\big( m_N,\hat\rho \big)$, representing upper bounds for $\hat\rho$ and corresponding lower bounds for $m_N$, have been included, in accordance with the CMS analysis of Ref.~\cite{CMShnl}.}
\end{figure}

Now we further discuss our aforementioned choice of values for the parameter $\hat\rho$, for which we refer the reader to the graph shown in Fig.~\ref{rhomasspars},
where the $CP$-even contribution $|\Delta\kappa|$, at a fixed CME of $\sqrt{s}=800\,{\rm GeV}$, has been plotted in the $\big( m_N,\hat\rho \big)$ plane. The $\hat\rho$ parameter has been taken to range within the interval $0\leqslant\hat\rho\leqslant1$, which corresponds to the vertical axis of this graph. On the other hand, values within $0\,{\rm GeV}\leqslant m_N\leqslant 1000\,{\rm GeV}$, along the horizontal axis, have been considered for the heavy-neutrino mass $m_N$. Darker tones in the graph represent smaller contributions, whereas the lighter the color tone the larger the contribution. Aiming at better appreciating the orders of magnitude within which this NP contribution ranges for the different values of $\big( m_N,\hat\rho \big)$, the $|\Delta\kappa|$ contribution has been plotted in base-10 logarithmic scale. The sizes of the contributions are specified by the labeling bar located beneath the graph. Lying in the upper-right section of the graph, a bounded region is displayed. It represents the set of $\big( m_N,\hat\rho \big)$ values which yield $|\Delta \kappa|$ contributions falling into expected ILC sensitivity~\cite{ILC2}. Within this ILC-sensitivity region, a few points have been plotted, which correspond to upper limits for $\hat\rho$ and lower bounds for $m_N$, in accordance with the exclusion regions provided by the CMS Collaboration in Ref.~\cite{CMShnl}. Thus, the region between these points and the boundary for ILC sensitivity comprehends the whole set of points producing $|\Delta\kappa|$ contributions consistent with both the CMS work and ILC expectations. Recall that the values $\hat\rho=0.58,0.65$ have been taken for our estimations and analyses. The points associated to these values have been plotted in the graph, from which one can appreciate that the corresponding lower limits for the heavy-neutrino mass are $m_N=650\,{\rm GeV}$ and $m_N=700\,{\rm GeV}$, respectively. We notice that both values $\hat\rho=0.58,0.65$ are well suited for $\sqrt{s}=800\,{\rm GeV}$, because a subset of the corresponding allowed heavy-neutrino masses yields some of the largest contributions to the AC $|\Delta\kappa|$.
\\

\begin{figure}[ht]
\center
\includegraphics[width=8cm]{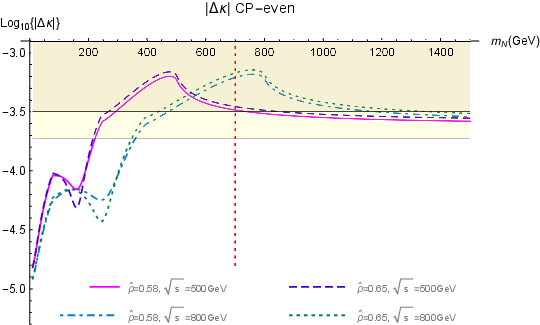}
\vspace{0.5cm}
\\
\includegraphics[width=8cm]{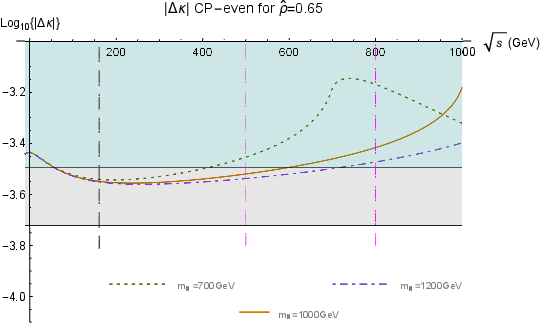}
\caption{\label{kappagraphs} One-loop AC contribution $|\Delta\kappa|$, in base-10 logarithmic scale, from virtual Majorana neutrinos. Upper graph: $\log_{10}\big\{|\Delta\kappa|\big\}$ as a function on the heavy-neutrino mass $m_N$, for fixed $\sqrt{s}$ and $\hat\rho$. Lower graph: $\log_{10}\big\{|\Delta\kappa|\big\}$ as a function on the CME $\sqrt{s}$, for fixed $m_N$ and $\hat\rho=0.65$. Both graphs include two horizontal lines, with the upper one corresponding to expected ILC sensitivity at $\sqrt{s}=500\,{\rm GeV}$ and the lower one representing ILC sensitivity at $\sqrt{s}=800\,{\rm GeV}$.}
\end{figure}

\begin{figure}[ht]
\center
\includegraphics[width=8cm]{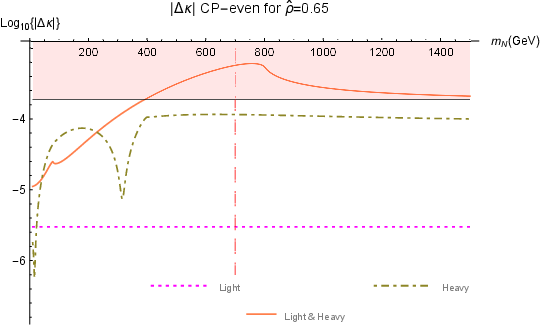}
\caption{\label{nuVSN} One-loop AC contribution $|\Delta\kappa|$, in base-10 logarithmic scale, from Feynman diagrams involving either light neutrinos (dotted curve), heavy neutrinos (dot-dashed curve), or both (solid curve). The contribution $\log_{10}\big\{|\Delta\kappa|\big\}$ is displayed as a function on the heavy-neutrino mass $m_N$, for $\sqrt{s}=800\,{\rm GeV}$ and $\hat\rho=0.65$. The graph includes a horizontal line, which corresponds to expected ILC sensitivity at $\sqrt{s}=800\,{\rm GeV}$.}
\end{figure}

Another depiction of the contributions $|\Delta\kappa|$ is provided by Fig.~\ref{kappagraphs}.
The upper panel of this figure displays plots of $|\Delta\kappa|$ as a function on the heavy-neutrino mass, for which $m_N$ values lying within $10\,{\rm GeV}\leqslant m_N\leqslant 1500\,{\rm GeV}$ have been considered. These curves have also been carried out in base-10 logarithmic scale. Two regions in the graph, which are bounded from below by horizontal solid lines corresponding to $|\Delta\kappa|=3.20\times10^{-4}$ and $|\Delta\kappa|=1.90\times10^{-4}$, represent the expected experimental sensitivity of ILC at $500\,{\rm GeV}$ (darkest region) and $800\,{\rm GeV}$ (lightest region), respectively, as estimated in Ref.~\cite{ILC2}. A vertical dashed line, at $m_N=700\,{\rm GeV}$, displays the minimum heavy-neutrino mass value, in accordance with our picks for the parameter $\hat\rho$ and the results of Ref.~\cite{CMShnl}. Four curves have been plotted, with each of them corresponding to the choices $\hat\rho=0.58,0.65$ and $\sqrt{s}=500\,{\rm GeV}, 800\,{\rm GeV}$, in the parameter space $\big( \hat\rho,\sqrt{s} \big)$. Around $m_N=700\,{\rm GeV}$, the long-dashed and the solid plots, both of them defined at the CME value $\sqrt{s}=500\,{\rm GeV}$, barely fall into the expected sensitivity region of ILC corresponding to such a CME. The curves given by $\sqrt{s}=800\,{\rm GeV}$, which are the dot-dashed and the dotted curves, fall within the ILC sensitivity capabilities for the heavy-neutrino mass range $700\,{\rm GeV}\leqslant m_N\lesssim1200\,{\rm GeV}$. In fact, both curves reach corresponding local maxima inside the ILC-sensitivity region at $\sqrt{s}=800\,{\rm GeV}$, near the heavy-neutrino-mass value $m_N=800\,{\rm GeV}$. Therefore, the observation of this effect would be plausible. The lower panel of Fig.~\ref{kappagraphs} shows the behavior of $|\Delta\kappa|$, in base-10 logarithmic scale, as a function on the CME $\sqrt{s}$, within the range $0\leqslant\sqrt{s}\leqslant1000\,{\rm GeV}$. To carry out these plots, the value $\hat\rho=0.65$ has been utilized. Furthermore, the three plots displayed in this graph are distinguished by the following heavy-neutrino mass values: $m_N=700\,{\rm GeV}$, from which the dotted curve follows; $m_N=1000\,{\rm GeV}$, yielding the solid plot; and the dot-dashed curve, given by $m_N=1200\,{\rm GeV}$. Just as in the previous graph, the sensitivity regions for ILC at $\sqrt{s}=500\,{\rm GeV}$ and $\sqrt{s}=800\,{\rm GeV}$, respectively bounded from below by the upper horizontal line and the lower horizontal line, have been included. This graph involves three vertical lines, one of which lies at $\sqrt{s}=2m_W$, thus representing the threshold for $W$-boson pair production. The other two vertical lines correspond to the CMEs $\sqrt{s}=500\,{\rm GeV}$ and $\sqrt{s}=800\,{\rm GeV}$. It can be appreciated, from this graph, that ILC sensitivity is slightly reached at $\sqrt{s}=500\,{\rm GeV}$ by the plot corresponding to $m_N=700\,{\rm GeV}$, whereas a measurement at a CME of $\sqrt{s}=800\,{\rm GeV}$ is more likely to be attainable, since the three plots reside within the ILC sensitivity region. Indeed notice, in this context, that the lowest allowed heavy-neutrino mass $m_N=700\,{\rm GeV}$ would be optimal in hope for a signal. The anomaly $\Delta\kappa$ receives contributions due to both light and heavy virtual neutrinos, which is illustrated by Fig.~\ref{nuVSN},
where we show a graph of $\log_{10}\big\{ |\Delta\kappa| \big\}$, as a function on the heavy-neutrino mass $m_N$, in GeV units, with $10\,{\rm GeV}\leqslant m_N\leqslant1500\,{\rm GeV}$. The horizontal solid line, which lower-bounds the colored region, represents the expected sensitivity achievable by ILC at a CME of $800\,{\rm GeV}$, whereas the vertical dashed line indicates the minimum $m_N$ value, at $700\,{\rm GeV}$. Three plots have been included in this graph, all of them carried out by taking $\hat\rho=0.65$ and a CME of $\sqrt{s}=800\,{\rm GeV}$. For starters, the horizontal short-dashed line represents the contributions exclusively associated to virtual light neutrinos. Meanwhile, the dot-dashed curve depicts the contributions produced by diagrams solely involving heavy-neutrino virtual lines. The third and last plot, which is solid, comes from contributing diagrams in which both heavy and light virtual neutrinos participate, through $Z\nu_j N_k$ vertices. With this in mind, note that, among these three contributions, the dominant effect, for most of the heavy-neutrino mass values within the $m_N$ range under consideration, is produced by the diagrams comprising heavy- and light-neutrino virtual lines. Such a contribution turns out to be larger than the smallest contribution, corresponding to virtual light neutrinos, by up to $\sim2$ orders of magnitude, while being not so far from the contributions from diagrams with heavy neutrinos only. Therefore, an observation of this effect should be understood as linked to NP heavy neutrinos. 
\\

\begin{figure}[ht]
\center
\includegraphics[width=8cm]{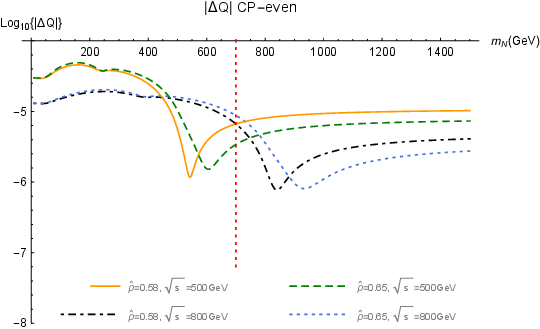}
\caption{\label{Qgraph} One-loop AC contribution $|\Delta Q|$, in base-10 logarithmic scale, from virtual Majorana neutrinos. We show $\log_{10}\big\{|\Delta Q|\big\}$ as a function on the heavy-neutrino mass $m_N$, for fixed $\sqrt{s}$ and $\hat\rho$.}
\end{figure}

Aiming at discussing the NP $CP$-preserving $\Delta Q$ contribution, we have carried out the graph shown in Fig.~\ref{Qgraph},
in which curves of $|\Delta Q|$, as determined by the heavy-neutrino mass $m_N$, have been executed, in base-10 logarithmic scale. The mass range which we considered is $10\,{\rm GeV}\leqslant m_N\leqslant1500\,{\rm GeV}$, with the vertical dotted line, at $m_N=700\,{\rm GeV}$, displaying the minimum allowed mass of heavy neutrino. All these curves, performed by usage of $\sqrt{s}=500\,{\rm GeV}, 800\,{\rm GeV}$ and $\hat\rho=0.58,0.65$, are well beyond the reach of the estimated sensitivity for ILC, by about 1 order of magnitude, so the corresponding ILC sensitivity regions have not been included in the graph.
\\

\subsection{CP odd contributions}
Next we address those contributions from Majorana neutrinos to $WWZ$ in which $CP$ symmetry is not preserved, where only $\Delta\tilde{\kappa}$ remains. The identification and exploration of sources of $CP$ violation is nowadays quite relevant indeed, since its occurrence is a necessary condition to explain the matter-antimatter asymmetry~\cite{Sakharov}. The only place in the SM where this phenomenon is generated is the complex phase in the Kobayashi-Maskawa mixing matrix~\cite{KoMa}, located in the quark sector. In this sense, the $CP$ violation is a well-motivated mean to look for NP, beyond the SM. While the SM has been found to lack $CP$-odd AC $\Delta\tilde\kappa_{\rm SM}$ and $\Delta\tilde Q_{\rm SM}$ at the one-loop level, NP models producing such effects exist. An instance of this is the investigation developed in Ref.~\cite{CPoddSusy}, whose authors profited from a $CP$-violation phase lying within mass matrices for neutralinos and charginos, which yields $CP$-violating interactions of the $W$ and $Z$ bosons with supersymmetric particles. According to that work, $CP$-odd contributions to $WWZ$ anomalous couplings as large as $10^{-3}$ might be generated. On the other hand, Ref.~\cite{AMOSS} has pointed out that the assumption of NP extra quarks, in the context of the vector-like quark model, introduces new sources of $CP$ non-conservation, thus being able to generate $CP$-odd contributions to $WWZ$ anomalies of order $10^{-5}$. The authors of Ref.~\cite{BuPi} explored the $CP$-odd contributions to $WWZ$ from Majorana neutrinos in a framework defined by the sequential introduction, to the SM, of a fourth lepton family. Contributions were found to be $\sim10^{-3}$. Technicolor theories were the framework considered by the authors of Ref.~\cite{AppWu} to explore BSM contributions to the $CP$-odd part of the $WWZ$ vertex.
\\

Following the structure of the general expression given in Eq.~(\ref{genanomaly}), we identify a coefficient $\Delta\tilde\kappa^{(1)}_{\alpha k}$ which is both mass and $s$ independent, and we change its notation as $\Delta\tilde\kappa^{(1)}_{\alpha k}\to\tilde\Omega$. Furthermore, we found that $\Delta\tilde\kappa^{(2)}_{\alpha kj}=\tilde{g}_{\alpha kj}+\tilde{\cal J}$, where $\tilde{\cal J}$ also does not depend neither on masses nor on $s$. The expressions of $\tilde\Omega$ and $\tilde{\cal J}$ are
\begin{equation}
\tilde\Omega=\frac{ig^2(2s_{\rm W}^2-1)}{(8\pi c_{\rm W})^2},\hspace{0.5cm} \tilde{\cal J}=\frac{2ig^2}{(16\pi c_{\rm W})^2}.
\label{tildesOmegaandJ}
\end{equation}
Regarding the contributing coefficient $\tilde g_{\alpha kj}$, we find it worth poninting out that it is antisymmetric with respect to its neutrino indices: $\tilde g_{\alpha kj}=-\tilde g_{\alpha jk}$. About the remaining contribution $\Delta\tilde\kappa^{(3)}_{\alpha kj}$, an alike antisymmetry property, $\Delta\tilde\kappa^{(3)}_{\alpha kj}=-\Delta\tilde\kappa^{(3)}_{\alpha jk}$, holds as well. The explicit expressions of the factors $\tilde g_{\alpha kj}$ and $\Delta\tilde\kappa^{(3)}_{\alpha kj}$ are
\begin{widetext}
\begin{eqnarray}
&&
\tilde{g}_{\alpha kj}=\frac{i g^2 }{(16 \pi m_W s c_{\rm W})^2}
\Big(-2m_W^2s\left(m_\alpha^2-m_W^2\right)\left(m_{n_k}^2-m_{n_j}^2\right) C_0^{(n_j,\alpha,n_k)}
\nonumber \\ && \hspace{1cm}
+2m_W^2s\left(m_\alpha^2-m_W^2\right)
\Big(
\Lambda_1^{(\alpha,n_k)}-\Lambda_1^{(\alpha,n_j)}
\Big)-2m_W^2s\left(m_{n_k}^2-m_{n_j}^2\right) \Lambda_2^{(n_k,n_j)}
\nonumber \\ && \hspace{1cm}
+s\left( m_\alpha^2-m_W^2 \right)
\Big(
-\left(m_\alpha^2-m_{n_k}^2-m_W^2\right)\log \Big(\frac{m_\alpha^2}{m_{n_k}^2}\Big)
+\left(m_\alpha^2-m_{n_j}^2-m_W^2\right)\log \Big(\frac{m_\alpha^2}{m_{n_j}^2}\Big)
\Big)
\nonumber \\ && \hspace{1cm}
+m_W^2\log \Big(\frac{m_{n_k}^2}{m_{n_j}^2}\Big) \left( \left( m_{n_k}^2-m_{n_j}^2 \right)^2-s\left( m_{n_k}^2+m_{n_j}^2 \right) \right)
\nonumber \\ && \hspace{1cm}
-2m_W^2s\left(m_{n_k}^2-m_{n_j}^2\right)\Big),
\end{eqnarray}
\begin{eqnarray}
&&
\Delta\tilde\kappa^{(3)}_{\alpha kj}=\frac{i g^2 m_{n_k} m_{n_j}}{s(16 \pi  m_W c_{\rm W})^2}
\Big(-2m_W^2\left(m_{n_k}^2-m_{n_j}^2\right) C_0^{(n_j,\alpha,n_k)}+2m_W^2
\Big(
\Lambda_1^{(\alpha,n_k)}-\Lambda_1^{(\alpha,n_j)}
\Big)
\nonumber \\ &&\hspace{1cm}
-\left(m_\alpha^2-m_{n_k}^2-m_W^2\right) \log \Big(\frac{m_\alpha^2}{m_{n_k}^2}\Big)+\left(m_\alpha^2-m_{n_j}^2-m_W^2\right) \log \Big(\frac{m_\alpha^2}{m_{n_j}^2}\Big)\Big).
\end{eqnarray}
\end{widetext}
The definitions of $C_0^{(n_j,\alpha,n_k)}$, $\Lambda_1^{(\alpha,j)}$, and $\Lambda_2^{(n_k,n_j)}$, all of them involved in these two equations, are provided in Appendix~\ref{App3}. Once the degenerate spectra for light- and heavy-neutrino masses are implemented, the factors $\Delta\tilde\kappa^{(3)}_{\alpha kj}$ split into $\Delta\tilde\kappa^{(3)}_{\alpha\nu\nu}$, $\Delta\tilde\kappa^{(3)}_{\alpha\nu N}$, $\Delta\tilde\kappa^{(3)}_{\alpha N\nu}$, and $\Delta\tilde\kappa^{(3)}_{\alpha NN}$, whereas, similarly, $\tilde{g}_{\alpha kj}$ unfolds into $\tilde{g}_{\alpha\nu\nu}$, $\tilde{g}_{\alpha\nu N}$, $\tilde{g}_{\alpha N\nu}$, and $\tilde{g}_{\alpha NN}$. However, the antisymmetry properties of these quantities imply that the coefficients $\Delta\tilde\kappa^{(3)}_{\alpha\nu\nu}$, $\Delta\tilde\kappa^{(3)}_{\alpha NN}$, $\tilde{g}_{\alpha\nu\nu}$, and $\tilde{g}_{\alpha NN}$ vanish. Meanwhile, the relations $\Delta\tilde\kappa^{(3)}_{\alpha\nu N}=-\Delta\tilde\kappa^{(3)}_{\alpha N\nu}$ and $\tilde{g}_{\alpha\nu N}=-\tilde{g}_{\alpha N\nu}$ are fulfilled. A further cancellation eliminates all the dependence on $\tilde{g}_{\alpha\nu N}$, so all what is left from $\Delta\tilde\kappa^{(2)}_{\alpha kj}$ contributions resides in $\tilde{\cal J}$, thus being mass independent. With all this in mind, we write down the total $CP$-odd contribution as $\Delta\tilde\kappa=\Delta\tilde\kappa_{\rm con.}+\Delta\tilde\kappa_{\rm par.}$, where
\begin{widetext}
\begin{equation}
\Delta\tilde\kappa_{\rm con.}=3\big( \tilde\Omega+\tilde{\cal J} \big),
\end{equation}
\begin{equation}
\Delta\tilde\kappa_{\rm par.}=\hat{\rho}^2{\rm tr}\Big\{ \big( {\bf 1}_3+\hat{\rho}^2XX^\dag \big)^{-\frac{1}{2}}
\Big( X\big( {\bf 1}_3+\hat{\rho}^2X^{\rm T}X^* \big)^{-1}X^{\rm T}-{\rm H.c.} \Big)\big( {\bf 1}_3+\hat{\rho}^2XX^\dag \big)^{-\frac{1}{2}}U_\nu^\dag\Delta\tilde\kappa^{(3)}_{\nu N}U_\nu \Big\},
\label{trDtildek}
\end{equation}
\end{widetext}
with $\tilde\kappa^{(3)}_{\nu N}$ a diagonal $3\times3$ matrix with entries $(\tilde\kappa^{(3)}_{\nu N})_{\alpha\beta}=\delta_{\alpha\beta}\,\tilde\kappa^{(3)}_{\beta\nu N}$. The term $\Delta\tilde\kappa_{\rm con.}$
is, according to Eq.~(\ref{tildesOmegaandJ}), constant with respect to all masses and $s$. Moreover, $\Delta\tilde\kappa_{\rm con.}$ does not depend on $\hat\rho$. By contrast, the term $\Delta\tilde\kappa_{\rm par.}$ bears dependence on these parameters, which is partly encoded in the matrix $\Delta\tilde\kappa^{(3)}_{\nu N}$, and also depends on the parameters defining the matrix $X$. If we implement the $X$-matrix texture given in Eq.~(\ref{Xtexture1}), the total contribution $\Delta\tilde\kappa$ reduces to
\begin{equation}
\Delta\tilde\kappa=3\big( \tilde\Omega+\tilde{\cal J} \big)+\frac{2i\hat{\rho}^2\sin2\varphi}{\big( 1+\hat{\rho}^2 \big)^2}\sum_\alpha\Delta\tilde\kappa^{(3)}_{\alpha\nu N}.
\end{equation}
Notice that the presence of a nonzero phase $\varphi$ is a necessary condition for the mass- and $s$-dependent term $\Delta\tilde\kappa_{\rm par.}$ to remain. This term has, however, a marginal impact, as it is smaller than the constant term $\Delta\tilde\kappa_{\rm con.}$ by more than 10 orders of magnitude. Such a noticeable difference comes from the contributing factors $\Delta\tilde\kappa_{\alpha\nu N}^{(3)}$, which are quite suppressed. Therefore, the $CP$-odd AC is, in practical terms, constant: $\Delta\tilde\kappa\approx\Delta\tilde\kappa_{\rm con.}$. With this in mind, we simply estimate the $CP$-odd contribution to be
\begin{equation}
|\Delta\tilde\kappa|=1.19\times10^{-3}.
\end{equation}
Recall that the ILC Technical Design Report has set the expectations on the ILC sensitivity to $CP$-odd AC. The corresponding estimations are displayed in Table~\ref{boundsbyILC}, from which we conclude that, while a relatively large $CP$-odd contribution $\Delta\tilde\kappa$ is generated, it would remain out of ILC sensitivity.
\\


\section{Summary}
\label{Concs}
The present work has been carried out aiming at an estimation of effects on the $WWZ$ vertex from neutrinos beyond the Standard Model, assumed to be described by Majorana fields. While the seesaw neutrino-mass-generating mechanism provides a nice explanation of tiny neutrino masses, which in this approach are determined by some high-energy scale connecting such masses with the large masses of a set of hypothetical heavy neutral leptons, current upper bounds on light-neutrino masses, which lie in the sub-eV range, dramatically push the high-energy scale towards grand-unification sized energy scales, thus avoiding direct production of new-physics particles and also largely suppressing contributions to Standard-Model observables. For the present investigation, we have considered a variant of the seesaw mechanism in which the strong link defined, in the framework of the ordinary seesaw, among light- and heavy-neutrino masses is weakened by leaving the light neutrinos massless at the tree level, then endowing them with radiatively-generated masses. This path allows for reasonably large values of heavy-neutrino masses. In this context, we addressed the one-loop contributions to the $WWZ$ vertex from virtual Majorana neutrinos, both light and heavy, running through loop lines in Feynman diagrams. Since no new-physics contributing one-loop diagrams with virtual gauge-boson lines are induced by this model, the calculation is gauge independent, even though the vertex has been calculated by taking the $Z$-boson external line to be off the mass shell. The reason for the assumed off-shellness of the vertex is that our study is carried out in the context of some future electron-positron collider, perhaps the International Linear Collider, able to produce $W$ boson pairs, a process which receives contributions from a $s$-channel diagram involving the $WWZ$ coupling, sensitive to the center-of-mass energy of the collision. In the presence of Majorana neutrinos, a larger number of contributing diagrams, in comparison to what we would expect from the Dirac case, is generated. These extra  contributions were found to coincide with those characterizing the Dirac case. The calculation was performed by following the tensor-reduction method, by Passarino and Veltman. We focused in the anomaly contributions $\Delta\kappa$, $\Delta Q$, $\Delta\tilde\kappa$, and $\Delta\tilde Q$, of which the first two are associated to new physics preserving $CP$ symmetry, whereas the remaining two are linked to $CP$ violation, a phenomenon which plays a role in the explanation of baryon asymmetry. After appropriate implementation of Schouten identities, we concluded that, at one loop, the $CP$-odd contribution $\Delta\tilde Q$ vanishes exactly, while $\Delta\tilde\kappa$ remained nonzero. All the non-vanishing contributions were found to be ultraviolet finite. The neutrino-mass model used for our phenomenological calculation requires the set of heavy-neutrino masses to be quasi-degenerate, so we took all of such masses to coincide with each other. Then, our analytical expressions were given in terms of three parameters, namely, the mass of the heavy neutrinos, the center-of-mass energy of the collision, and a parameter, $\hat\rho$, linked to light-heavy neutrino mixing. Based on previous papers on the matter, we considered the values $\hat\rho=0.58,0.65$ for our estimations. Note that such choices establish a lower limit $\sim700\,{\rm GeV}$, on the heavy neutrino mass, in accordance with results by the CMS Collaboration, at the CERN. Our estimations for the $CP$-even anomalies $\Delta\kappa$ and $\Delta Q$ were then presented and discussed. Since $\Delta Q$ turned out to be well below the expected sensitivity of the International Linear Collider, we centered our attention in $\Delta\kappa$. We found that this new-physics contribution might be within future experimental sensitivity for a $e^+e^-$ collider operating at a CME of $800\,{\rm GeV}$, and that any measurement would be due to the effects from virtual heavy neutrinos. Contributions as large as $\sim 10^{-3}$ were estimated. Then we turned our attention to the $CP$-odd contribution $\Delta\tilde\kappa$, which is absent in the SM, where it can be generated only since the three-loop level. Our estimations yield a contribution which amounts to $\sim10^{-3}$. This contribution lies around one order of magnitude beyond expected sensitivity for the International Linear Collider. For the sake of completeness, we also report the following estimations for other AC, present in the $WWZ$ vertex function given by Eqs.~(\ref{Gammaeven}) and (\ref{Gammaodd}): $|f_1|=7.68\times10^{-3}$, $|\tilde{f}_1|=1.69\times10^{-13}$, and $|\tilde{f}_2|=1.16\times10^{-8}$. These values have been determined by taking $\sqrt{s}=800\,{\rm GeV}$, $m_N=700\,{\rm GeV}$, and $\hat\rho=0.65$.

\section*{Acknowledgements}
\noindent
We acknowledge financial support from Conahcyt (M\'exico). M. S. acknowledges funding by Conahcyt, though the program ``Estancias Posdoctorales por M\'exico 2022.''

\appendix

\section{The $WWZ$ parametrization}
\label{App2}
Under the assumption of ${\rm U}(1)_e$ invariance, the $WWZ$ effective Lagrangian ${\cal L}^{WWZ}_{\rm eff}={\cal L}_{WWZ}^{\rm even}+{\cal L}^{\rm odd}_{WWZ}$ is constructed. Here, ${\cal L}^{\rm even}_{WWZ}$, comprised exclusively by $CP$-invariant terms, and ${\cal L}_{WWZ}^{\rm odd}$, not abiding by symmetry with respect to such a discrete transformation, have the following explicit expressions~\cite{EllWu,DiNa}:
\begin{eqnarray}
&&
{\cal L}_{WWZ}^{\rm even}=-ig_Z
\Big(
g_1\big( W^+_{\mu\nu}W^{-\mu} -W^-_{\mu\nu}W^{+\mu} \big)Z^\nu
\nonumber \\ &&
\hspace{0.6cm}
+\kappa\,W^+_\mu W^-_\nu Z^{\mu\nu}+\frac{\lambda}{m_W^2}W^+_{\mu\nu}W^{-\nu}\hspace{0.00001cm}_\rho Z^{\rho\mu}
\nonumber \\ &&
\hspace{0.6cm}
-ig_2\epsilon^{\rho\mu\lambda\nu}Z_\rho\big( W^+_\nu\partial_\lambda W^-_\mu-W^-_\mu\partial_\lambda W^+_\nu \big)
\Big),
\label{LCPeven}
\end{eqnarray}
\begin{eqnarray}
&&
{\cal L}_{WWZ}^{\rm odd}=-ig_Z
\Big(
\tilde{\kappa}W^+_\mu W^-_\nu\tilde{Z}^{\mu\nu}+\frac{\tilde{\lambda}}{m_W^2}W^+_{\mu\nu}W^{-\nu}\hspace{0.000001cm}_\rho\tilde{Z}^{\rho\mu}
\nonumber \\ &&
\hspace{1.3cm}
-i\tilde{g}_1W^+_\mu W^-_\nu\big( \partial^\mu Z^\nu+\partial^\nu Z^\mu \big)
\Big).
\label{LCPodd}
\end{eqnarray}
For these equations to be written down, the Lorentz 2-tensors $W^+_{\mu\nu}=\partial_\mu W^+_\nu-\partial_\nu W^+_\mu$, $W^-_{\mu\nu}=\partial_\mu W^-_\nu-\partial_\nu W^-_\mu$, and $Z_{\mu\nu}=\partial_\mu Z_\nu-\partial_\nu Z_\mu$ have been defined. These expressions also involve the dual tensor $\tilde{Z}^{\mu\nu}=\frac{1}{2}\epsilon^{\mu\nu\alpha\beta}Z_{\alpha\beta}$. The effective-Lagrangian terms of Eqs.~(\ref{LCPeven}) and (\ref{LCPodd}) yield, in accordance with the conventions displayed in Fig~\ref{WWZ}, the vertex function $\Gamma^{WWZ}_{\sigma\rho\mu}=\Gamma^{\rm even}_{\sigma\rho\mu}+\Gamma^{\rm odd}_{\sigma\rho\mu}$, given by Eqs.~(\ref{Gammaeven}) and (\ref{Gammaodd}), which are displayed above, in the Introduction. For the connection among the effective-Lagrangian terms and the vertex functions to consistently happen, the relations 
$\Delta\kappa=-g_1+\kappa+\lambda$, $\Delta Q=-2\lambda$, $g_2=-f_1$, $\Delta\tilde{\kappa}=\tilde{\kappa}+\frac{m_W^2-2q^2}{m_W^2}\tilde{\lambda}$, $\Delta\tilde{Q}=-2\tilde{\lambda}$, $2\tilde{g_1}=\tilde{f}_1$, $\tilde{f}_2=-\frac{4\tilde{\lambda}}{m_W^2}$, must hold. 
\\


\section{The neutrino model}
\label{App1}
Throughout the present Appendix, we discuss the neutrino-mass model given in Ref.~\cite{Pilaftsis}, which is the framework for our phenomenological calculation. 
\\

Assume that a theory beyond the Standard Model (BSM) is subjected to a couple of stages of spontaneous symmetry breaking, first at some high-energy scale $\Lambda$ and then at $v=246\,{\rm GeV}$, after which the corresponding Lagrangian, gauge invariant under ${\rm U}(1)_e$, is written as
\begin{equation}
{\cal L}_{\rm BSM}={\cal L}^\nu_{\rm mass}+{\cal L}_{W\nu l}+{\cal L}_{Z\nu\nu}+{\cal L}_{h\nu\nu}+\cdots.
\label{LBSM}
\end{equation}
The neutrino-mass Lagrangian,  ${\cal L}^\nu_{\rm mass}$, given by
\begin{eqnarray}
&&
{\cal L}^\nu_{\rm mass}=-\sum_{j=1}^3\sum_{k=1}^3\Big( \overline{\nu^0_{j,L}}(m_{\rm D})_{jk}\nu^0_{k,R}
\nonumber \\ &&
\hspace{1.2cm}
+\frac{1}{2}\overline{\nu^{0{\rm c}}_{j,R}}(m_{\rm M})_{jk}\nu^0_{k,R} \Big)+{\rm H.c.},
\end{eqnarray}
comprises three left-handed neutrino fields $\nu^0_{j,L}$ and three right-handed neutrino fields $\nu^0_{j,R}$, together with the charge-conjugated spinor fields $\nu^{0{\rm c}}_{j,R}$. The $3\times3$ lepton-number-violating matrix $m_{\rm M}$ is assumed to originate form the symmetry breaking at $\Lambda$, whereas the Dirac-like matrix $m_{\rm D}$, $3\times3$-sized, is assumed to come from electroweak symmetry breaking. Then, ${\cal L}_{\rm mass}^\nu$ is rearranged as
\begin{equation}
{\cal L}^\nu_{\rm mass}=-\frac{1}{2}\big( \overline{f_L}\,\,\,\,\,\overline{F_L} \big){\cal M}
\left(
\begin{array}{c}
f_R
\vspace{0.2cm}
\\
F_R
\end{array}
\right)+{\rm H.c.},
\end{equation}
where $(f_L)_j=\nu^0_{j,L}$, $(F_L)_j=\nu_{j,R}^{0{\rm c}}$, $(f_R)_j=\nu_{j,L}^{0{\rm c}}$, and $(F_R)_j=\nu^0_{j,R}$. Moreover, 
\begin{equation}
{\cal M}=\left(
\begin{array}{cc}
0 & m_{\rm D}
\vspace{0.2cm}
\\
m_{\rm D}^{\rm T} & m_{\rm M}
\end{array}
\right).
\end{equation}
Since $m_{\rm M}$ is symmetric, the property ${\cal M}^{\rm T}={\cal M}$ holds, which guarantees the existence of a diagonalization unitary matrix~\cite{Takagi}
\begin{equation}
{\cal U}=
\left(
\begin{array}{cc}
{\cal U}_{11} & {\cal U}_{12}
\vspace{0.2cm}
\\
{\cal U}_{21} & {\cal U}_{22}
\end{array}
\right),
\label{Ublocks}
\end{equation}
here written in terms of $3\times3$ block matrices ${\cal U}_{jk}$. 
In this context, the diagonalization operates as
\begin{equation}
{\cal U}^{\rm T}{\cal M}\,\,{\cal U}=
\left(
\begin{array}{cc}
M_\nu & 0
\vspace{0.2cm}
\\
0 & M_N
\end{array}
\right),
\label{numassdiagonalization}
\end{equation}
where $M_\nu$ and $M_N$ are $3\times3$ diagonal real matrices, whose eigenvalues are, respectively, the three-level light- and heavy-neutrino masses. The diagonalization procedure induces a change of basis, which defines three massless neutrino fields $\nu_j$ and three heavy-neutrino fields $N_j$, all of them fulfilling the Majorana condition. In Ref.~\cite{Pilaftsis}, the conditions $({\cal M}{\cal U})_{jk}=0$, for $j=1,2,3,4,5,6$, are introduced as sufficient and necessary for the $k$-th light neutrino to be massless at tree level. In this manner, the mass terms for light neutrinos are eliminated from ${\cal L}^\nu_{\rm mass}$, though keep in mind that such assumptions do not mess with heavy-neutrino masses, which remain nonzero at tree level. By this mean, the tie originally defined by the seesaw mechanism to link the masses of light neutrinos to the masses of their heavy counterparts has been broken. Light neutrinos, known to be massive in accordance with the occurrence of neutrino oscillations, do get massive, but their masses follow from radiative corrections. Ref.~\cite{Pilaftsis} provides a calculation of one-loop masses of light neutrinos, from which a remarkable statement is made: a new link between light- and heavy-neutrino masses is established, according to which the former are tiny enough as long as the heavy-neutrino mass spectrum is quasi-degenerate.  
\\

From the block-matrix form of the unitary diagonalization matrix ${\cal U}$, as given in Eq.~(\ref{Ublocks}), the following quantities are defined:
\begin{equation}
{\cal B}_{\alpha \nu_j}=\sum_{k=1}^3V^\ell_{\alpha k}\big( {\cal U}^*_{11} \big)_{kj},
\label{Bnu}
\end{equation}
\begin{equation}
{\cal B}_{\alpha N_j}=\sum_{k=1}^3V^\ell_{\alpha k}\big( {\cal U}^*_{12} \big)_{kj}.
\label{BN}
\end{equation}
Here, $\alpha=e,\mu,\tau$, that is, this greek index labels the flavors of the SM leptons. The $3\times3$ matrix $V^\ell$ is analogous to the Kobayashi-Maskawa quark-mixing matrix~\cite{KoMa}, but for the lepton sector. The set of quantities ${\cal B}_{\alpha\nu_j}$ and ${\cal B}_{\alpha N_j}$, just defined in Eqs.~(\ref{Bnu}) and (\ref{BN}), respectively define the $3\times3$ matrices ${\cal B}_\nu$ and ${\cal B}_N$, which can be accommodated as matrix blocks into the $3\times6$ matrix 
\begin{equation}
{\cal B}=
\big(
\begin{array}{cc}
{\cal B}_\nu & {\cal B}_N
\end{array}
\big).
\label{Binblocks} 
\end{equation}
The entries of ${\cal B}$ are generically denoted as ${\cal B}_{\alpha j}$, in which case the relations
\begin{equation}
{\cal B}_{\alpha j}=
\left\{
\begin{array}{lcl}
{\cal B}_{\alpha \nu_k}\textrm{, if } j=1,2,3,
\vspace{0.3cm}
\\
{\cal B}_{\alpha N_k}\textrm{, if } j=4,5,6,
\end{array}
\right.
\end{equation}
with $\nu_k=\nu_1,\nu_2,\nu_3$ and $N_k=N_1,N_2,N_3$, hold. The matrix ${\cal B}$ satisfies a sort of one-sided unitarity property: 
\begin{eqnarray}
{\cal B}{\cal B}^\dag={\bf 1}_3,&\textrm{ or }&\sum_{j=1}^6{\cal B}_{\alpha j}{\cal B}^*_{\beta j}=\delta_{\alpha\beta},
\end{eqnarray}
\begin{eqnarray}
{\cal B}^\dag{\cal B}={\cal C},&\textrm{ or }&\sum_{\alpha=e,\mu,\tau}{\cal B}^*_{\alpha j}{\cal B}_{\alpha k}={\cal C}_{jk}.
\end{eqnarray}
In these equations, ${\bf 1}_3$ has been used to denote the $3\times3$ identity matrix. Furthermore, the quantities ${\cal C}_{jk}$ are the entries of a $6\times6$ Hermitian matrix ${\cal C}$, which we express in terms of $3\times3$ matrix blocks as
\begin{equation}
{\cal C}=
\left(
\begin{array}{cc}
{\cal C}_{\nu\nu} & {\cal C}_{\nu N}
\vspace{0.2cm}
\\
{\cal C}_{N\nu} & {\cal C}_{NN}
\end{array}
\right).
\label{Cinblocks}
\end{equation}
The entries of ${\cal C}$ are related to those of these matrix blocks as
\begin{equation}
{\cal C}_{jk}=
\left\{
\begin{array}{ll}
{\cal C}_{\nu_i\nu_l},&\textrm{ if } j=1,2,3 \textrm{ and } k=1,2,3,
\vspace{0.2cm}
\\
{\cal C}_{\nu_iN_l},&\textrm{ if } j=1,2,3 \textrm{ and } k=4,5,6,
\vspace{0.2cm}
\\
{\cal C}_{N_i\nu_l},&\textrm{ if }  j=4,5,6 \textrm{ and } k=1,2,3,
\vspace{0.2cm}
\\
{\cal C}_{N_iN_l},&\textrm{ if }  j=4,5,6 \textrm{ and } k=4,5,6,
\end{array}
\right.
\end{equation}
where $\nu_i,\nu_l=\nu_1,\nu_2,\nu_3$ and $N_i,N_l=N_1,N_2,N_3$. Moreover, the entries of the matrix blocks are defined, in terms of the unitary diagonalization matrix given in Eq.~(\ref{Ublocks}), as
\begin{equation}
{\cal C}_{\nu_i\nu_l}=\sum_{j=1}^3\big( {\cal U}_{11} \big)_{ji}\big( {\cal U}^*_{11} \big)_{jl},
\end{equation}
\begin{equation}
{\cal C}_{\nu_iN_l}=\sum_{j=1}^3\big( {\cal U}_{11} \big)_{ji}\big( {\cal U}^*_{12} \big)_{jl},
\end{equation}
\begin{equation}
{\cal C}_{N_i\nu_l}=\sum_{j=1}^3\big( {\cal U}_{12} \big)_{ji}\big( {\cal U}^*_{11} \big)_{jl},
\end{equation}
\begin{equation}
{\cal C}_{N_iN_l}=\sum_{j=1}^3\big( {\cal U}_{12} \big)_{ji}\big( {\cal U}^*_{12} \big)_{jl}.
\end{equation}
The matrix ${\cal C}$ fulfills
\begin{eqnarray}
{\cal C}{\cal C}^\dag={\cal C},&\textrm{ or }&\sum_{i=1}^6{\cal C}_{ji}{\cal C}^*_{ki}={\cal C}_{jk}.
\end{eqnarray}
\\

Returning to the Lagrangian ${\cal L}_{\rm BSM}$, as expressed in Eq.~(\ref{LBSM}), the explicit expressions of the Lagrangian terms ${\cal L}_{W\nu l}$, ${\cal L}_{Z\nu\nu}$, and ${\cal L}_{h\nu\nu}$ read
\begin{eqnarray}
&&
{\cal L}_{W\nu l}=\sum_{\alpha=e,\mu,\tau}\sum_{j=1}^3\frac{g}{\sqrt{2}}\Big(
{\cal B}_{\alpha\nu_j}W^-_\rho\overline{l_\alpha}\gamma^\rho P_L\nu_j
\nonumber \\ &&
\hspace{1.5cm}
+{\cal B}_{\alpha N_j}W^-_\rho\overline{l_\alpha}\gamma^\rho P_LN_j
\Big)+{\rm H.c.},
\label{LCC}
\end{eqnarray}
\begin{eqnarray}
&&
{\cal L}_{Z\nu\nu}=\sum_{k=1}^3\sum_{j=1}^3
\frac{g}{4c_{\rm W}}\Big(
Z_\rho\overline{\nu_k}\gamma^\rho\big( i{\cal C}^{\rm Im}_{\nu_k\nu_j}-{\cal C}^{\rm Re}_{\nu_k\nu_j}\gamma_5 \big)\nu_j
\nonumber \\ &&
\hspace{0.8cm}
+\Big( Z_\rho\overline{\nu_k}\gamma^\rho\big( i{\cal C}^{\rm Im}_{\nu_kN_j}-{\cal C}^{\rm Re}_{\nu_kN_j}\gamma_5 \big)N_j+{\rm H.c.} \Big)
\nonumber \\ &&
\hspace{0.8cm}
+Z_\rho\overline{N_k}\gamma^\rho\big( i{\cal C}^{\rm Im}_{N_kN_j}-{\cal C}^{\rm Re}_{N_kN_j}\gamma_5 \big)N_j
\Big),
\label{LNC}
\end{eqnarray}
\begin{eqnarray}
&&
{\cal L}_{h\nu\nu}=\sum_{k=1}^3\sum_{j=1}^3
\frac{g}{4m_W}\Big(
h\overline{\nu_k}
\big(
(m_{\nu_k}+m_{\nu_j}){\cal C}^{\rm Re}_{\nu_k\nu_j}
\nonumber \\
&&
\hspace{2cm}
-i\gamma_5(m_{\nu_k}-m_{\nu_j}){\cal C}^{\rm Im}_{\nu_k\nu_j}
\big)\nu_j
\nonumber \\
&&
\hspace{2cm}
+h\overline{\nu_k}
\big(
(m_{\nu_k}+m_{N_j}){\cal C}^{\rm Re}_{\nu_k N_j}
\nonumber \\ 
&&
\hspace{2cm}
-i\gamma_5(m_{\nu_k}-m_{N_j}){\cal C}^{\rm Im}_{\nu_kN_j}
\big)N_j
\nonumber \\ 
&&
\hspace{2cm}
+h\overline{N_k}
\big(
(m_{N_k}+m_{N_j}){\cal C}^{\rm Re}_{N_kN_j}
\nonumber \\ &&
\hspace{2cm}
-i\gamma_5(m_{N_k}-m_{N_j}){\cal C}^{\rm Im}_{N_kN_j}
\big)N_j
\Big).
\end{eqnarray}
In these equations, $g$ is the ${\rm SU}(2)_L$ coupling constant, $c_{\rm W}$ is a short notation for $\cos\theta_{\rm W}$, and $P_L=\frac{1}{2}({\bf 1}_4-\gamma_5)$ is the left chiral projection matrix. Moreover, $W_\rho$ denotes the SM $W$-boson field, whereas $Z_\rho$ does it for the SM $Z$-boson field, $h$ is the Higgs field, and $l_\alpha$ stands for the SM $\alpha$-flavor charged lepton. To write down these couplings, the matrix ${\cal C}$ has been expressed as ${\cal C}={\cal C}^{\rm Re}+i{\cal C}^{\rm Im}$, with ${\cal C}^{\rm Re}={\rm Re}\{ {\cal C} \}$ and ${\cal C}^{\rm Im}={\rm Im}\{ {\cal C} \}$, which also applies for the matrix blocks comprising ${\cal C}$, shown in Eq.~(\ref{Cinblocks}).
\\

The $6\times6$ unitary matrix ${\cal U}$, which diagonalizes the mass matrix ${\cal M}$ as displayed in Eq.~(\ref{numassdiagonalization}), can be block-parametrized, in terms of some $3\times3$ complex matrix $\xi$, as~\cite{KPS,DePi}
\begin{equation}
{\cal U}=
\left(
\begin{array}{cc}
\big( {\bf 1}_3+\xi^*\xi^{\rm T} \big)^{-\frac{1}{2}} & \xi^*\big( {\bf 1}_3+\xi^{\rm T}\xi^* \big)^{-\frac{1}{2}}
\vspace{0.2cm}
\\
-\xi^{\rm T}\big( {\bf 1}_3+\xi^*\xi^{\rm T} \big)^{-\frac{1}{2}} & \big( {\bf 1}_3+\xi^{\rm T}\xi^* \big)^{-\frac{1}{2}}
\end{array}
\right).
\label{Uapprox}
\end{equation}
Using this parametrization, the matrices ${\cal B}$ and ${\cal C}$, respectively given in Eqs.~(\ref{Binblocks}) and (\ref{Cinblocks}), can be written in terms of $\xi$:
\begin{equation}
{\cal B}=
\Big(
\begin{array}{ccc}
V^\ell\big( {\bf 1}_3+\xi\xi^\dag \big)^{-\frac{1}{2}}
&&
V^\ell\xi\big( {\bf 1}_3+\xi^\dag\xi \big)^{-\frac{1}{2}}
\end{array}
\Big),
\label{Binxi}
\end{equation}
\begin{equation}
{\cal C}=
\left(
\begin{array}{cc}
\big( {\bf 1}_3+\xi\xi^\dag\big)^{-1} & \big( {\bf 1}_3+\xi\xi^\dag \big)^{-1}\xi
\vspace{0.2cm}
\\
\xi^\dag\big( {\bf 1}_3+\xi\xi^\dag \big)^{-1} & \xi^\dag\big( {\bf 1}_3+\xi\xi^\dag \big)^{-1}\xi
\end{array}
\right).
\label{Cinxi}
\end{equation}
If the assumption that the moduli of all the $\xi$ entries, $|\xi_{jk}|$, are small is taken, the matrix $\xi$ reduces to 
\begin{equation}
\xi=m_{\rm D}m_{\rm M}^{-1}.
\label{xidef}
\end{equation}
Moreover, while due to the afore-imposed condition for tree-level light-neutrino mass cancellation we have $M_\nu=0$, the diagonal heavy-neutrino mass matrix is given by $M_N\simeq m_{\rm M}\big( {\bf 1}_3+\frac{1}{2}m_{\rm M}^{-1}(\xi^\dag m_{\rm D}+m_{\rm D}^{\rm T}\xi^*) \big)$~\cite{Pilaftsis}.
\\


\section{The $\Delta\zeta^{(a)}$ factors}
\label{App3}
In this Appendix we provide explicit expressions of factors $\Delta\zeta^{(a)}$, which determine the anomaly $CP$-even contributions $\Delta\kappa$ and $\Delta Q$, in accordance with Eqs.~(\ref{Dkdegenerate})-(\ref{DQdegenerate}). Note that all light-neutrino masses have been taken to be equal to each other, so $m_{\nu_j}=m_\nu$ for $j=1,2,3$, whereas, similarly, all the masses of heavy neutrinos have been taken the same, that is, $m_{N_j}=m_N$ for $j=1,2,3$. In this context, we generically label light- and heavy- neutrinos stuff by the sole index $n$, thus meaning that, for instance, $m_n=m_\nu,m_N$. Corresponding to any contribution $\Delta\zeta^{(2)}$ or $\Delta\zeta^{(3)}$ we have, for any fixed lepton flavor $\alpha$, three different factors: $\Delta\zeta^{(a)}_{\alpha\nu\nu}$, $\Delta\zeta^{(a)}_{\alpha\nu N}$, and $\Delta\zeta^{(a)}_{\alpha NN}$. Keep in mind that $\Delta\zeta^{(a)}_{\alpha\nu\nu}$ and $\Delta\zeta^{(a)}_{\alpha NN}$ share the same structure, only differing by their neutrino-mass dependence, which is given by either $m_\nu$ or $m_N$. Among the three factors, on the other hand, $\Delta\zeta^{(a)}_{\alpha\nu N}$ is the one with the most general structure, which indeed particularizes to that of $\Delta\zeta^{(a)}_{\alpha\nu\nu}$ or $\Delta\zeta^{(a)}_{\alpha NN}$ if we take $m_N=m_\nu$. Also let us comment that the explicit expressions of $\Delta\zeta^{(a)}_{\alpha\nu N}$, for both $\Delta\kappa$ and $\Delta Q$, are very large, so we opted for not showing them. Instead, we considered more reasonable to solely display the definitions of $\Delta\zeta^{(a)}_{\alpha nn}$. These expressions involve the Passarino-Veltman 3-point scalar function $C_0$, defined in Eq.~(\ref{PaVeC0}), as well as the disc function~\cite{Patel}
\begin{eqnarray}
&&
\Lambda\big( p^2,m_0^2,m_1^2 \big)=\frac{\sqrt{\lambda\big( p^2,m_0^2,m_1^2 \big)}}{p^2}
\nonumber \\ &&\hspace{0.5cm}
\times
\log\Big(\textstyle \frac{2m_0m_1}{-p^2+m_0^2+m_1^2-\sqrt{\lambda\big( p^2,m_0^2,m_1^2 \big)}}+i\epsilon \Big),
\end{eqnarray}
with $\lambda(a,b,c)=a^2+b^2+c^2$ the K\"allen function. For the sake of succinctness, we use the following notation:
\begin{equation}
\Lambda_1^{(A,B)}=\Lambda\big( m^2_W,m^2_A,m^2_B \big),
\end{equation}
\begin{equation}
\Lambda_2^{(A)}=\Lambda\big( s,m^2_A,m^2_A \big),
\end{equation}
\begin{equation}
C_0^{(A,B,C)}=C_0(m_W^2,m_W^2,s,m_A^2,m_B^2,m_C^2).
\end{equation}
Then, we write down the $\Delta\zeta^{(a)}$ factors as
\begin{eqnarray}
&&
\Delta\kappa^{(1)}_{\alpha n}=\frac{g^2}{D^{(1)}}
\Big(
\eta^{(1)}_{1,\alpha n}+\eta^{(1)}_{2,\alpha n}\log\Big( \frac{m_\alpha^2}{m_n^2} \Big)+\eta^{(1)}_{3,\alpha n}\Lambda_1^{(\alpha,n)}
\nonumber \\ && \hspace{1cm}
+\eta^{(1)}_{4,\alpha n}\Lambda_2^{(\alpha)}+\eta^{(1)}_{5,\alpha n}C_0^{(\alpha,n,\alpha)}
\Big),
\end{eqnarray}
\begin{eqnarray}
&&
\Delta\kappa^{(2)}_{\alpha nn}=\frac{g^2}{D^{(2)}}
\Big(
\eta^{(2)}_{1,\alpha n}+\eta^{(2)}_{2,\alpha n}\log\Big( \frac{m_\alpha^2}{m_n^2} \Big)+\eta^{(2)}_{3,\alpha n}\Lambda_1^{(\alpha,n)}
\nonumber \\ && \hspace{1cm}
+\eta^{(2)}_{4,\alpha n}\Lambda_2^{(n)}+\eta^{(2)}_{5,\alpha n}C_0^{(n,\alpha,n)}
\Big),
\end{eqnarray}
\begin{eqnarray}
&&
\Delta\kappa^{(3)}_{\alpha nn}=\frac{g^2}{D^{(3)}}
\Big(
\eta^{(3)}_{1,\alpha n}\log\Big( \frac{m_\alpha^2}{m_n^2} \Big)+\eta^{(3)}_{2,\alpha n}\Lambda_1^{(\alpha,n)}
\nonumber \\ && \hspace{1cm}
+\eta^{(3)}_{3,\alpha n}\Lambda_2^{(n)}+\eta^{(3)}_{4,\alpha n}C_0^{(n,\alpha,n)}
\Big),
\end{eqnarray}
\begin{eqnarray}
&&
\Delta Q^{(1)}_{\alpha n}=\frac{g^2}{\bar{D}^{(1)}}\Big(
\sigma^{(1)}_{1,\alpha n}+\sigma^{(1)}_{2,\alpha n}\log\Big( \frac{m_\alpha^2}{m_n^2} \Big)+\sigma^{(1)}_{3,\alpha n}\Lambda_1^{(\alpha,n)}
\nonumber \\ && \hspace{1cm}
+\sigma^{(1)}_{4,\alpha n}\Lambda_2^{(\alpha)}+\sigma^{(1)}_{5,\alpha n}C_0^{(\alpha,n,\alpha)}
\Big),
\end{eqnarray}
\begin{eqnarray}
&&
\Delta Q^{(2)}_{\alpha nn}=\frac{g^2}{\bar{D}^{(2)}}
\Big(
\sigma^{(2)}_{1,\alpha n}+\sigma^{(2)}_{2,\alpha n}\log\Big( \frac{m_\alpha^2}{m_n^2} \Big)+\sigma^{(2)}_{3,\alpha n}\Lambda_1^{(\alpha,n)}
\nonumber \\ && \hspace{1cm}
+\sigma^{(2)}_{4,\alpha n}\Lambda_2^{(n)}+\sigma^{(2)}_{5,\alpha n}C_0^{(n,\alpha,n)}
\Big),
\end{eqnarray}
where
\begin{equation}
D^{(1)}=3\big( 8\pi c_{\rm W} \big)^2\big( 4m_W^2-s \big)^3,
\end{equation}
\begin{equation}
D^{(2)}=3\big( 16\pi s\,m_Wc_{\rm W} \big)^2\big( 4m_W^2-s \big)^3,
\end{equation}
\begin{equation}
D^{(3)}=\big( 16\pi m_W c_{\rm W} \big)^2\big( 4m_W^2-s \big) s,
\end{equation}
\begin{equation}
\bar{D}^{(1)}=\frac{3\big( 4\pi c_{\rm W} \big)^2\big( 4m_W^2-s \big)^3s}{2s_{\rm W}^2-1},
\end{equation}
\begin{equation}
\bar{D}^{(2)}=3\big( 8\pi c_{\rm W} \big)^2\big( s(4m_W^2-s) \big)^3.
\end{equation}
Furthermore, the coefficients $\eta^{(a)}$ and $\sigma^{(a)}$, in terms of which these contributions have been written, are given by
\begin{widetext}
\begin{eqnarray}
&&
\eta^{(1)}_{1,\alpha n}=\frac{2}{m_W^2} \left(2 s_{\rm W}^2-1\right) \left(4m_W^2-s\right) \Big(8 m_W^6+3 s m_W^4+10 s m_n^2 m_W^2+\left(24 m_W^2-s\right) m_{\alpha}^4-2 \big(\left(24 m_W^2-s\right) m_n^2
\nonumber \\ && \hspace{1cm}
+m_W^2 \left(16 m_W^2+s\right)\big) m_{\alpha}^2+m_n^4 \left(24 m_W^2-s\right)\Big),
\end{eqnarray}
\begin{eqnarray}
&&
\eta^{(1)}_{2,\alpha n}=-\frac{1}{m_W^4}
\Big(
\left(2 s_{\rm W}^2-1\right) \left(144m_W^4-10 s m_W^2+s^2\right) m_{\alpha }^6+\big(6 m_W^2 \big(-8 \left(8s_{\rm W}^2-5\right) m_W^4
\nonumber \\ &&\hspace{1cm}
+s \left(5-18 s_{\rm W}^2\right) m_W^2+s^2 s_{\rm W}^2\big)-3\left(2 s_{\rm W}^2-1\right) m_n^2 \left(144 m_W^4-10 s m_W^2+s^2\right)\big) m_{\alpha}^4+3 \big(32 \left(s_{\rm W}^2-1\right) m_W^8
\nonumber \\ &&\hspace{1cm}
+2 s \left(30 s_{\rm W}^2-11\right)m_W^6-2 s^2 s_{\rm W}^2 m_W^4+\left(2 s_{\rm W}^2-1\right) m_n^4 \left(144 m_W^4-10 sm_W^2+s^2\right)+m_n^2 \big(16 \left(8 s_{\rm W}^2-5\right) m_W^6
\nonumber \\ &&\hspace{1cm}
+4 s \left(22s_{\rm W}^2-9\right) m_W^4-s^2 m_W^2\big)\big) m_{\alpha }^2-\left(2s_{\rm W}^2-1\right) \big(\left(144 m_W^4-10 s m_W^2+s^2\right) m_n^6+3 s m_W^2 \left(26m_W^2+s\right) m_n^4
\nonumber \\ &&\hspace{1cm}
+3 \left(16 m_W^8+6 s m_W^6+5 s^2 m_W^4\right) m_n^2+s m_W^6 \left(26m_W^2+s\right)\big)
\Big),
\end{eqnarray}
\begin{eqnarray}
&&
\eta^{(1)}_{3,\alpha n}=-\frac{2}{m_W^2}\Big(-\left(2 s_{\rm W}^2-1\right) \left(144 m_W^4-10s m_W^2+s^2\right) m_{\alpha }^4+\big(96 \left(s_{\rm W}^2-1\right) m_W^6
\nonumber \\ &&\hspace{1cm}
+8 s \left(16s_{\rm W}^2-5\right) m_W^4+s^2 \left(1-8 s_{\rm W}^2\right) m_W^2+2 \left(2s_{\rm W}^2-1\right) m_n^2 \left(144 m_W^4-10 s m_W^2+s^2\right)\big) m_{\alpha}^2
\nonumber \\ &&\hspace{1cm}
-\left(2 s_{\rm W}^2-1\right) \left(\left(144 m_W^4-10 s m_W^2+s^2\right) m_n^4+4m_W^2 \left(12 m_W^4+8 s m_W^2+s^2\right) m_n^2+s m_W^4 \left(26m_W^2+s\right)\right)\Big),
\nonumber \\ 
\end{eqnarray}
\begin{eqnarray}
&&
\eta^{(1)}_{4,\alpha n}=-2\Big(6 \left(2 s_{\rm W}^2-1\right)\left(8 m_W^2+3 s\right) m_{\alpha }^4+\big(32 \left(s_{\rm W}^2+1\right) m_W^4+4 s\left(13-38 s_{\rm W}^2\right) m_W^2+6 s^2 s_{\rm W}^2
\nonumber \\ &&\hspace{1cm}
-12 \left(2 s_{\rm W}^2-1\right)m_n^2 \left(8 m_W^2+3 s\right)\big) m_{\alpha }^2+\left(2 s_{\rm W}^2-1\right) \big(6\left(8 m_W^2+3 s\right) m_n^4+3 \left(16 m_W^4+4 s m_W^2+3 s^2\right) m_n^2
\nonumber \\ &&\hspace{1cm}
+s m_W^2\left(26 m_W^2+s\right)\big)\Big),
\end{eqnarray}
\begin{eqnarray}
&&
\eta^{(1)}_{5,\alpha n}=6\Big(2 \left(2 s_{\rm W}^2-1\right) \left(8 m_W^2+3 s\right) m_{\alpha}^6+\big(16 m_W^4+s \left(22-60 s_{\rm W}^2\right) m_W^2+s^2
\nonumber \\ &&\hspace{1cm}
-6 \left(2s_{\rm W}^2-1\right) m_n^2 \left(8 m_W^2+3 s\right)\big) m_{\alpha }^4+\big(-32s_{\rm W}^2 m_W^6+2 s \left(26 s_{\rm W}^2-9\right) m_W^4+s^2 \left(4s_{\rm W}^2-3\right) m_W^2
\nonumber \\ &&\hspace{1cm}
+6 \left(2 s_{\rm W}^2-1\right) m_n^4 \left(8 m_W^2+3s\right)+m_n^2 \left(-16 m_W^4+4 s \left(18 s_{\rm W}^2-7\right) m_W^2+s^2 \left(12s_{\rm W}^2-7\right)\right)\big) m_{\alpha }^2
\nonumber \\ &&\hspace{1cm}
-\left(2 s_{\rm W}^2-1\right) \big(2\left(8 m_W^2+3 s\right) m_n^6+6 s \left(m_W^2+s\right) m_n^4+\left(-16 m_W^6+18 sm_W^4+s^3\right) m_n^2
\nonumber \\ &&\hspace{1cm}
+2 s m_W^4 \left(m_W^2+s\right)\big)\Big),
\end{eqnarray}
\begin{eqnarray}
&&
\eta^{(2)}_{1,\alpha n}=8 s^2 \left(4 m_W^2-s\right)\big(m_\alpha^4 \left(24 m_W^2-s\right)+2 m_\alpha^2\left(m_n^2 \left(s-24 m_W^2\right)+5 m_W^2s\right)
\nonumber \\ &&\hspace{1cm}
+\left(m_n^2-m_W^2\right) \big(m_n^2 \left(24m_W^2-s\right)-8 m_W^4-3 m_W^2 s\big)\big),
\end{eqnarray}
\begin{eqnarray}
&&
\eta^{(2)}_{2,\alpha n}=-\frac{2}{m_W^2}\Big( 2 s^2\big(m_\alpha^6\left(144 m_W^4-10 m_W^2 s+s^2\right)+m_\alpha^4 \left(3m_W^2 s \left(26 m_W^2+s\right)-3 m_n^2 \left(144 m_W^4-10m_W^2 s+s^2\right)\right)
\nonumber \\ &&\hspace{1cm}
+3 m_\alpha^2 \left(m_n^4 \left(144m_W^4-10 m_W^2 s+s^2\right)-m_n^2 m_W^2 \left(80 m_W^4+36m_W^2 s+s^2\right)+16 m_W^8+6 m_W^6 s+5 m_W^4s^2\right)
\nonumber \\ &&\hspace{1cm}
-m_n^6 \left(144 m_W^4-10 m_W^2 s+s^2\right)+30 m_n^4m_W^4 \left(8 m_W^2+s\right)-6 m_n^2 \left(16 m_W^8+11m_W^6 s\right)+m_W^6 s \left(26 m_W^2+s\right)\big) \Big),
\nonumber \\
\end{eqnarray}
\begin{eqnarray}
&&
\eta^{(2)}_{3,\alpha n}=8 s^2 \Big(m_\alpha^4 \left(144 m_W^4-10 m_W^2 s+s^2\right)+m_\alpha^2 \left(4m_W^2 \left(12 m_W^4+8 m_W^2 s+s^2\right)-2 m_n^2 \left(144m_W^4-10 m_W^2 s+s^2\right)\right)
\nonumber \\ &&\hspace{1cm}
+m_n^4 \left(144 m_W^4-10m_W^2 s+s^2\right)+m_n^2 \left(-96 m_W^6-40 m_W^4 s+m_W^2s^2\right)+m_W^4 s \left(26 m_W^2+s\right)\Big),
\end{eqnarray}
\begin{eqnarray}
&&
\eta^{(2)}_{4,\alpha n}=-8m_W^2 s^2 \Big(6 m_\alpha^4 \left(8 m_W^2+3s\right)+m_\alpha^2 \left(-12 m_n^2 \left(8 m_W^2+3 s\right)+48m_W^4+12 m_W^2 s+9 s^2\right)+6 m_n^4 \left(8 m_W^2+3 s\right)
\nonumber \\ &&\hspace{1cm}
-4m_n^2 \left(8 m_W^4+13 m_W^2 s\right)+m_W^2 s \left(26m_W^2+s\right)\Big),
\end{eqnarray}
\begin{eqnarray}
&&
\eta^{(2)}_{5,\alpha n}=-24 m_W^2 s^2 \Big(2 m_\alpha^6 \left(8 m_W^2+3 s\right)-6m_\alpha^4 \left(m_n^2 \left(8 m_W^2+3 s\right)-s\left(m_W^2+s\right)\right)+m_\alpha^2 \big(6 m_n^4 \left(8m_W^2+3 s\right)
\nonumber \\ &&\hspace{1cm}
-m_n^2 \left(16 m_W^4+28 m_W^2 s+7 s^2\right)-16m_W^6+18 m_W^4 s+s^3\big)-\left(m_n^2-m_W^2\right) \big(2m_n^4 \left(8 m_W^2+3 s\right)
\nonumber \\ &&\hspace{1cm}
-m_n^2 s \left(16 m_W^2+s\right)+2m_W^2 s \left(m_W^2+s\right)\big)\Big),
\end{eqnarray}
\begin{equation}
\eta^{(3)}_{1,\alpha n}=4m_n^2 s \left(m_\alpha^2-m_n^2+m_W^2\right),
\end{equation}
\begin{equation}
\eta^{(3)}_{2,\alpha n}=-8m_n^2 m_W^2 s,
\end{equation}
\begin{equation}
\eta^{(3)}_{3,\alpha n}=8m_n^2 m_W^2 s,
\end{equation}
\begin{equation}
\eta^{(3)}_{4,\alpha n}=8m_n^2 m_W^2 s \left(m_\alpha^2-m_n^2+m_W^2\right),
\end{equation}
\begin{eqnarray}
&&
\sigma^{(1)}_{1,\alpha n}=\left(4 m_W^2-s\right) \Big(2 \left(6 m_W^2+s\right) m_{\alpha}^4-4 \left(\left(6 m_W^2+s\right) m_n^2+2 m_W^2 \left(m_W^2+s\right)\right) m_{\alpha}^2+s^2 m_W^2
\nonumber \\ &&\hspace{1cm}
+12 \left(m_W^3-m_n^2 m_W\right){}^2+2 s \left(m_n^4+8 m_W^2m_n^2-m_W^4\right)\Big)
\end{eqnarray}
\begin{eqnarray}
&&
\sigma^{(1)}_{2,\alpha n}=\frac{1}{m_W^2}\Big(\left(2 m_W^6-3 \left(m_{\alpha }^2-5 m_n^2\right) m_W^4+6 m_n^2\left(m_n-m_{\alpha }\right) \left(m_n+m_{\alpha }\right) m_W^2+\left(m_{\alpha}^2-m_n^2\right){}^3\right) s^2
\nonumber \\ &&\hspace{1cm}
+2 m_W^2\left(5 m_W^6-6 \left(2 m_n^2+3 m_{\alpha}^2\right) m_W^4+3 \left(5 m_n^4-12 m_{\alpha }^2 m_n^2+7 m_{\alpha }^4\right) m_W^2-8\left(m_{\alpha }^2-m_n^2\right){}^3\right) s
\nonumber \\ &&\hspace{1cm}
-12 m_W^4 \left((m_W-m_{\alpha })^2-m_n^2\right) \left((m_W+m_{\alpha })^2-m_n^2\right) \left(-m_n^2+m_W^2+m_{\alpha }^2\right)\Big),
\end{eqnarray}
\begin{eqnarray}
&&
\sigma^{(1)}_{3,\alpha n}=\Big(24\left(\left(m_{\alpha }^2-m_n^2\right){}^2-m_W^4\right) m_W^4+4 s \big(5 m_W^4+\left(5m_n^2-13 m_{\alpha }^2\right) m_W^2
\nonumber \\ &&\hspace{1cm}
+8 \left(m_{\alpha }^2-m_n^2\right){}^2\big) m_W^2-2s^2 \left(-2 m_W^4+\left(m_{\alpha }^2-5 m_n^2\right) m_W^2+\left(m_{\alpha}^2-m_n^2\right){}^2\right)\Big),
\end{eqnarray}
\begin{eqnarray}
&&
\sigma^{(1)}_{4,\alpha n}=2m_W^2\Big(6 \left(2 m_W^2-3 s\right) m_{\alpha}^4+\left(-8 m_W^4+20 s m_W^2+3 s^2+m_n^2 \left(36 s-24 m_W^2\right)\right) m_{\alpha}^2
\nonumber \\ &&\hspace{1cm}
+12 \left(m_W^3-m_n^2 m_W\right){}^2-s^2 \left(9 m_n^2+2 m_W^2\right)-2 s \left(9m_n^4-6 m_W^2 m_n^2+5 m_W^4\right)\Big),
\end{eqnarray}
\begin{eqnarray}
&&
\sigma^{(1)}_{5,\alpha n}=-6m_W^2\Big(\left(4 m_W^2-6 s\right) m_{\alpha }^6+2 \left(-2 m_W^4+4s m_W^2+s^2+m_n^2 \left(9 s-6 m_W^2\right)\right) m_{\alpha }^4
\nonumber \\ &&\hspace{1cm}
+2 \left(-2 m_W^6+sm_W^4-2 s^2 m_W^2+m_n^4 \left(6 m_W^2-9 s\right)+m_n^2 \left(-4 m_W^4+2 s m_W^2-4s^2\right)\right) m_{\alpha }^2+4 m_W^2 \left(m_W^2-m_n^2\right){}^3
\nonumber \\ &&\hspace{1cm}
+s^3 m_n^2+2 s^2\left(3 m_n^4-m_W^2 m_n^2+m_W^4\right)+2 s \left(3 m_n^6-6 m_W^2 m_n^4+5 m_W^4 m_n^2-2m_W^6\right)\Big),
\end{eqnarray}
\begin{eqnarray}
&&
\sigma^{(2)}_{1,\alpha n}=-4s^2 \left(4 m_W^2-s\right) \Big(2 m_\alpha^4 \left(6m_W^2+s\right)-4 m_\alpha^2 \left(m_n^2 \left(6m_W^2+s\right)+6 m_W^4-4 m_W^2 s\right)+2 m_n^4 \left(6m_W^2+s\right)
\nonumber \\ &&\hspace{1cm}
-8 m_n^2 m_W^2 \left(m_W^2+s\right)+m_W^2\left(12 m_W^4-2 m_W^2 s+s^2\right)\Big),
\end{eqnarray}
\begin{eqnarray}
&&
\sigma^{(2)}_{2,\alpha n}=-\frac{4s^2}{m_W^2} \Big(m_\alpha^6\left(-12 m_W^4-16 m_W^2 s+s^2\right)+m_\alpha^4 \big(m_n^2\left(36 m_W^4+48 m_W^2 s-3 s^2\right)+6 m_W^2 \big(6 m_W^4
\nonumber \\ &&\hspace{1cm}
-5m_W^2 s-s^2\big)\big)-3 m_\alpha^2 \big(m_n^4 \left(12m_W^4+16 m_W^2 s-s^2\right)+m_n^2 \left(8 m_W^6-24 m_W^4s-2 m_W^2 s^2\right)+12 m_W^8
\nonumber \\ &&\hspace{1cm}
-8 m_W^6 s+5 m_W^4s^2\big)+\left(m_n^2-m_W^2\right)^2 \left(m_n^2 \left(12m_W^4+16 m_W^2 s-s^2\right)+2 m_W^2 \left(6 m_W^4-5 m_W^2s-s^2\right)\right)\Big),
\nonumber \\ 
\end{eqnarray}
\begin{eqnarray}
&&
\sigma^{(2)}_{3,\alpha n}=8s^2 \Big(m_\alpha^4 \left(-12 m_W^4-16m_W^2 s+s^2\right)+m_\alpha^2 \left(m_n^2 \left(24 m_W^4+32m_W^2 s-2 s^2\right)-5 m_W^2 s \left(2m_W^2+s\right)\right)
\nonumber \\ && \hspace{1cm}
-\left(m_n^2-m_W^2\right) \left(m_n^2\left(12 m_W^4+16 m_W^2 s-s^2\right)+2 m_W^2 \left(6 m_W^4-5m_W^2 s-s^2\right)\right)\Big),
\end{eqnarray}
\begin{eqnarray}
&&
\sigma^{(2)}_{4,\alpha n}=-8m_W^2 s^2 \Big(6 m_\alpha^4 \left(2 m_W^2-3s\right)-3 m_\alpha^2 \left(4 m_n^2 \left(2 m_W^2-3 s\right)+8m_W^4-4 m_W^2 s+3 s^2\right)+6 m_n^4 \left(2 m_W^2-3s\right)
\nonumber \\ &&\hspace{1cm}
+m_n^2 \left(-8 m_W^4+20 m_W^2 s+3 s^2\right)+2 m_W^2\left(6 m_W^4-5 m_W^2 s-s^2\right)\Big),
\end{eqnarray}
\begin{eqnarray}
&&
\sigma^{(2)}_{5,\alpha n}=24m_W^2 s^2 \Big(m_\alpha^6 \left(6 s-4 m_W^2\right)+6m_\alpha^4 \left(m_n^2 \left(2 m_W^2-3 s\right)+2 m_W^4-2m_W^2 s+s^2\right)+m_\alpha^2 \big(-6 m_n^4 \left(2 m_W^2-3s\right)
\nonumber \\ &&\hspace{1cm}
+m_n^2 \left(-8 m_W^4+4 m_W^2 s-8 s^2\right)-12 m_W^6+10m_W^4 s-2 m_W^2 s^2+s^3\big)+2 \left(m_n^2-m_W^2\right)^2\big(m_n^2 \left(2 m_W^2-3 s\right)
\nonumber \\ &&\hspace{1cm}
+2 m_W^4-2 m_W^2s+s^2\big)\Big).
\end{eqnarray}
\end{widetext}


\begin{thebibliography}{99}
%
\bibitem{Kamiokande} Y. Fukuda {\it et al}. (Super-Kamiokande Collaboration), Evidence for Oscillation of Atmospheric Neutrinos, Phys. Rev. Lett. {\bf 81}, 1562 (1998).
%
\bibitem{SNO} Q. R. Ahmad {\it et al}. (SNO Collaboration), Direct Evidence for Neutrino Flavor Transformation from Neutral-Current Interactions in the Sudbury Neutrino Observatory, Phys. Rev. Lett. {\bf 89}, 011301 (2002).
%
\bibitem{Glashow} [1] S. L. Glashow, Partial-symmetries of weak interactions, Nucl. Phys. {\bf 22}, 579 (1961).
%
\bibitem{Salam} A. Salam, Weak and electromagnetic interactions, Conf. Proc. C {\bf 680519}, 367 (1968).
%
\bibitem{Weinberg} S. Weinberg, A Model of Leptons, Phys. Rev. Lett. {\bf 19}, 1264 (1967).
%
\bibitem{Pontecorvo} B. Pontecorvo, Mesonium and anti-mesonium, Sov. Phys. JETP {\bf 6}, 429 (1957).
%
\bibitem{Dirac} P. A. M. Dirac, The quantum theory Proc. R. Soc. A {\bf 117}, 610 (1928).
%
\bibitem{MoSe1} R. N. Mohapatra and G. Senjanovi\'c, Neutrino Mass and Spontaneous Parity Nonconservation, Phys. Rev. Lett. {\bf 44},
912 (1980).
%
\bibitem{MoSe2} R. N. Mohapatra and G. Senjanovi\'c, Neutrino masses and mixings in gauge models with spontaneous parity violation, Phys. Rev. D {\bf 23}, 165 (1981).
%
\bibitem{PaSa} J.C. Pati and A. Salam, Lepton number as the fourth``color'', Phys. Rev. D {\bf 10}, 275 (1974).
%
\bibitem{Majorana} E. Majorana, Teoria simmetrica dell?elettrone e del positrone, Nuovo Cimento {\bf 14}, 171 (1937).
%
\bibitem{Weinbergoperator} S. Weinberg, Baryon- and Lepton-Nonconserving Processes, Phys. Rev. Lett. {\bf 43}, 1566 (1979).
%
\bibitem{DGMP} A. Dobado, A. G\'omez-Nicola, A. L. Maroto, and J. R. Pel\'aez, Effective Lagrangians for the Standard Model (Springer-Verlag, Berlin/Heidelberg, 1997). 
%
\bibitem{EnBr} F. Englert and R. Brout, Broken Symmetry and the Mass of Gauge Vector Mesons, Phys. Rev. Lett. {\bf 13}, 321 (1964).
%
\bibitem{Higgs} P. W. Higgs, Broken Symmetries and the Masses of Gauge Bosons, Phys. Rev. Lett. {\bf 13}, 508 (1964).
%
\bibitem{KATRIN} M. Aker {\it et al}. (The KATRIN Collaboration), Direct neutrino-mass measurement with sub-electronvolt sensitivity, Nature Phys. {\bf 18}, 160 (2022).
%
\bibitem{SDSSexperiment} S. Alam {\it et al}., Completed SDSS-IV extended Baryon Oscillation Spectroscopic Survey: Cosmological implications from two decades of spectroscopic surveys at the Apache Point Observatory, Phys. Rev. D {\bf 103}, 083533 (2021).
%
\bibitem{Planckexperiment} N. Aghanim {\it et al}. (Planck Collaboration), Planck 2018 results. VL. Cosmological parameters, Astron. Astrophys. {\bf 641}, A6 (2020).
%
\bibitem{CHLR} Y. Cai, T. Han, T. Li, and R. Ruiz, Lepton Number Violation: Seesaw Models and Their Collider Tests, Front. Phys. {\bf 6}, 40 (2018).
%
\bibitem{MoVa} R. N. Mohapatra and J. W. F. Valle, Neutrino mass and baryon-number nonconservation in superstring models, Phys. Rev. D {\bf 34}, 1642 (1986).
%
\bibitem{GoVa} M. C. Gonzalez-Garcia and J. W. F. Valle, Fast decaying neutrinos and observable flavour violation in a new class of majoron models, Phys. Lett. B {\bf 216}, 360 (1989).
%
\bibitem{DeVa} F. Deppisch and J. W. F. Valle, Enhanced lepton flavor violation in the supersymmetric inverse seesaw model, Phys. Rev. D {\bf 72}, 036001 (2005). 
%
\bibitem{ALSV1} E. Akhmedov, M. Lindner, E. Schnapka, and J. W. F. Valle, Left-right symmetry breaking in NJL approach, Phys. Lett. B {\bf 368}, 270 (1996).
%
\bibitem{ALSV2} E. Akhmedov, M. Lindner, E. Schnapka, and J. W. F. Valle, Dynamical left-right symmetry breaking, Phys. Rev. D {\bf 53}, 2752 (1996).
%
\bibitem{Pilaftsis} A. Pilaftsis, Radiatively induced neutrino masses and large Higgs-neutrino couplings
in the Standard Model with Majorana fields, Z. Phys. C {\bf 55}, 275 (1992).
%
\bibitem{Feynman} R. P. Feynman, Space-Time Approach to Quantum Electrodynamics, Phys. Rev. {\bf 76}, 769 (1949).
%
\bibitem{BGL} W.A. Bardeen, R. Gastmans, and B. Lautrup, Static quantities in Weinberg's model of weak and electromagnetic interactions, Nucl. Phys. {\bf B46}, 319 (1972).
%
\bibitem{HPZH} K. Hagiwara, R. D. Peccei, D. Zeppenfeld, and K. Hikasa, Probing the weak boson sector in $e^+e^-\to W^+W^-$, Nucl. Phys. {\bf B282}, 253 (1987).
%
\bibitem{BaZe} U. Baur and D. Zeppenfeld, Probing the $WW\gamma$ vertex at future colliders, Nucl. Phys. {\bf B308}, 127 (1988).
%
\bibitem{AKLPS} E. N. Argyres, G. Katsilieris, A. B. Lahanas, C. G. Papadopoulos, and V. C. Spanos, One-loop corrections to three-vector-boson vertices in the Standard Model, Nucl. Phys, {\bf B391}, 23 (1993).
%
\bibitem{PaPhi} J. Papavassiliou and K. Philippides, Gauge-invariant three-boson vertices in the standard model and the static properties of the $W$, Phys. Rev. D {\bf 48}, 4255 (1993).
%
\bibitem{CKL} D. Chang, W. -Y. Keung, and J.Liu, The electric dipole moment of the $W$-boson, Nucl. Phys. {\bf B355}, 295 (1991).
%
\bibitem{PoKhr} M. Pospelov and I. Khriplovich, Electric dipole moment of the W boson and the electron in the Kobayashi-Maskawa model, Sov. J. Nucl.
Phys. {\bf 53}, 638 (1991).
%
\bibitem{BoGi} C. G. Bollini and J. J. Giambiagi, Dimensional renormalization: The number of dimensions as a regularizing parameter, Nuovo Cimento Soc. Ital. Fis. {\bf 12B}, 20 (1972).
%
\bibitem{tHooVe} G. 't Hooft and M. Veltman, Regularization and renormalization of gauge fields, Nucl. Phys. {\bf B44}, 189 (1972).
%
\bibitem{DEHK} A. Denner, H. Eck, O. Hahn, and J. K\"ublbeck, Feynman rules for fermion-number-violating interactions, Nucl. Phys. {\bf B387}, 467 (1992).
%
\bibitem{GlZr} J. Gluza and M. Zraek, Feynman rules for Majorana-neutrino interactions, Phys. Rev. D {\bf 45}, 1693  (1992).
%
\bibitem{ILC1} G. Weiglein {\it et al}. (The LHC/ILC Study Group), Physics interplay of the LHC and the ILC, Phys. Rep. {\bf 426}, 47 (2006).
%
\bibitem{ILC2} H. Baer {\it et al}. (ILC Collaboration), The international linear collider technical design report-Volume 2: Physics, arXiv:1306.6352.
%
\bibitem{MMNS} E. Mart\'inez, J. Monta\~no-Dom\'inguez, H. Novales-S\'anchez, and M. Salinas, New physics in $WW\gamma$ at one loop via Majorana neutrinos, Phys. Rev. D {\bf 107}, 035025 (2023).
%
\bibitem{NoSa} H. Novales-S\'anchez and M. Salinas, Majorana neutrinos in the triple gauge boson coupling $ZZZ^*$, Phys. Rev. D {\bf 108}, 075032 (2023).
%
\bibitem{LLR} C. N. Leung, S. T. Love, and S. Rao, Low-energy manifestations of a new interactions scale: Operator analysis, Z. Phys. C {\bf 31}, 433 (1986).
%
\bibitem{BuWy} W. Buchm\"uller and D. Wyler, Effective lagrangian analysis of new interactions and flavour conservation, Nucl. Phys. {\bf B268}, 621 (1986).
%
\bibitem{Wudka} J. Wudka, Electroweak effective Lagrangians, Int. J. Mod. Phys. A {\bf 09}, 2301 (1994).
%
\bibitem{EllWu} J. Ellison and J. Wudka, Study of trilinear gauge-boson couplings at the Tevatron collider, Ann. Rev. Nucl. Part. Sci. {\bf 48}, 33 (1998).
%
\bibitem{Ward} J. C. Ward, An identity in quantum electrodynamics, Phys. Rev. {\bf 78}, 182 (1950).
%
\bibitem{BRS1} C. Becchi, A. Rouet, and R. Stora, Renormalization of the abelian Higgs-Kibble model, Commun. Math. Phys. 42, 127 (1975).
%
\bibitem{BRS2} C. Becchi, A. Rouet, and R. Stora, Renormalization of gauge theoriesAnn. Phys. (N.Y.) 98, 287 (1976).
%
\bibitem{Tyutin} I. V. Tyutin, Gauge Invariance in Field Theory and Statistical Physics in Operator Formalism, FIAN (P.N: Lebedev Physical Institute of the USSR Academy of Science), Report No. 39, 1975.
%
\bibitem{GPS} J. Gomis, J. Paris, and S. Samuel, Antibracket, antifields and gauge-theory quantization, Phys. Rep. {\bf 259}, 1 (1995).
%
\bibitem{Taylor} J. C. Taylor, Ward identities and charge renormalization of the Yang-Mills field, Nucl. Phys. {\bf B33}, 436 (1971).
%
\bibitem{Slavnov} A. A. Slavnov, Ward identities in gauge theories, Theor. Math. Phys. {\bf 10}, 99 (1972).
%
\bibitem{BSH} M. B\"ohm, H. Spiesberger, and W. Hollik, On the One Loop Renormalization of the Electroweak Standard Model and Its Application to Leptonic Processes, Fortsch. Phys. {\bf 34}, 687 (1986).
%
\bibitem{DeWitt} B.S. DeWitt, Quantum Theory of Gravity. II. The Manifestly Covariant Theory, Phys. Rev. {\bf 162}, 1195 (1967).
%
\bibitem{tHooftBFM} G. 't Hooft, The Background Field Method in Gauge Field Theories, Acta Univ. Wratislavensis {\bf 368}, 345 (1976).
%
\bibitem{Abbott1} L. F. Abbott, The Background Field Method Beyond One Loop, Nucl. Phys. {\bf B185}, 189 (1981).
%
\bibitem{Abbott2} L. F. Abbott, Introduction to the Background Field Method, Acta Phys. Pol. B {\bf 13}, 33 (1982).
%
\bibitem{Cornwall} J. M. Cornwall, Dynamical mass generation in continuum quantum chromodynamics, Phys. Rev. D {\bf 26}, 1453 (1982).
%
\bibitem{CoPa} J. M. Cornwall and J. Papavassiliou, Gauge-invariant three-gluon vertex in QCD, Phys. Rev. D {\bf 40}, 3474 (1989).
%
\bibitem{Papavassiliou} J. Papavassiliou, Gauge-invariant proper self-energies and vertices in gauge theories with broken symmetry, Phys. Rev.
D {\bf 41}, 3179 (1990).
%
\bibitem{DDW} A. Denner, S. Dittmaier, and G. Weiglein, Gauge invariance, gauge parameter independence and properties of Green functions, in Proceedings of
the Ringberg Workshop ``Perspectives for Electroweak Interactions in $e^+e^-$ Collisions'' Ringberg, Germany, 1995, edited by B. A. Kniehl (World Scientific, Singapore, 1995), p. 281, Report No. hep-ph/9505271 (unpublished). 
%
\bibitem{PapavassiliouRef} J. Papavassiliou, Standard model higher order corrections to the $WW\gamma/WWZ$ vertex, AIP Conf. Proc. {\bf 350}, 98 (1995).
%
\bibitem{PDG} R.L. Workman {\it et al}. (Particle Data Group), Review of Particle Physics, Prog. Theor. Exp. Phys. 2022, 083C01 (2022).
%
\bibitem{CUPIDMOndbd} E. Armengaud {\it et al}. (CUPID-Mo Collaboration), New limit for neutrinoless double-beta decay of $^{100}$Mo from the CUPID-Mo experiment, Phys. Rev. Lett. {\bf 126}, 181802 (2021).
%
\bibitem{CUOREndbd} D.Q. Adams {\it et al}. (CUORE Collaboration), Improved limit on neutrinoless double-beta decay in $^{130}$Te with CUORE, Phys. Rev. Lett. {\bf 124}, 122501 (2020).
%
\bibitem{GERDAndbd} M. Agostini {\it et al}. (GERDA Collaboration), Final results of GERDA on the search for neutrinoless double-$\beta$ decay, Phys. Rev. Lett. {\bf 125}, 252502 (2020).
%
\bibitem{Majoranandbd} S.I. Alvis {\it et al}. (Majorana Collaboration), Search for neutrinoless double-$\beta$ decay in $^{76}$Ge with 26 kg yr of exposure from the Majorana demonstrator, Phys. Rev. C {\bf 100}, 025501 (2019).
%
\bibitem{EXO200ndbd} G. Anton {\it et al}. (EXO-200 Collaboration), Search for neutrinoless double-$\beta$ decay with the complete EXO-200 dataset, Phys. Rev. Lett. {\bf 123}, 161802 (2019).
%
\bibitem{KamLANDZenndbd} A. Gando {\it et al}. (KamLAND-Zen Collaboration), Search for Majorana neutrinos near the inverted mass hierarchy region with KamLAND-zen, Phys. Rev. Lett. {\bf 117}, 082503 (2016).
%
\bibitem{nohayndbd} A. Gando {\it et al}. (KamLAND-Zen Collaboration), Limit on neutrinoless $\beta\beta$ decay of 1$^{36}$Xe from the first phase of KamLAND-Zen and comparison with the positive claim in $^{76}$Ge, Phys. Rev. Lett. {\bf 110}, 062502 (2013).
%
\bibitem{MRS} J. M. M\'arquez, P. Roig, and M. Salinas, $\nu e\to\nu e$ scattering with massive Dirac or Majorana neutrinos and general interactions, arXiv:2401.14305 [hep-ph].
%
\bibitem{MNSmatrix} Z. Maki, M. Nakagawa, and S. Sakata, Remarks on the unified model of elementary particles, Prog. Theor. Phys. {\bf 28}, 870 (1962).
%
\bibitem{Pontecorvomatrix} B. Pontecorvo, Neutrino experiments and the problem of conservation of leptonic charge, Sov. Phys. JETP {\bf 26}, 984 (1968).
%
\bibitem{BGS} C. Broggini, C. Giunti, and S. Studenikin, Electromagnetic Properties of Neutrinos, Adv. High Energy Phys. {\bf 2012}, 459526 (2012).
%
\bibitem{PaVe} G. Passarino and M. Veltman, One-loop corrections for $e^+e^-$ annihilation into $\mu^+\mu^-$ in the Weinberg model, Nucl. Phys. {\bf B160}, 151 (1979).
%
\bibitem{DeSt} G. Devaraj and R. G. Stuart, Reduction of one-loop tensor form factors to scalar integrals: A general scheme, Nucl. Phys. {\bf B519}, 483 (1998).
%
\bibitem{SMO1} V. Shtabovenko, R. Mertig, and F. Orellana, \textsc{FeynCalc} 9.3: New features and improvements, Comput. Phys. Commun. {\bf 256}, 107478 (2020).
%
\bibitem{SMO2} V. Shtabovenko, R. Mertig, and F. Orellana, New developments in \textsc{FeynCalc} 9.0, Comput. Phys. Commun. {\bf 207}, 432 (2016).
%
\bibitem{MBD} R. Mertig, M. Bohm, and A. Denner, \textsc{FeynCalc}--Computer-algebraic calculation of Feynman amplitudes, Comput. Phys. Commun. {\bf 64}, 345 (1991).
%
\bibitem{Patel} H. H. Patel, \textsc{Package-X}: A Mathematica package for the analytic calculation of one-loop integrals, Comput. Phys. Commun. {\bf 197}, 276 (2015).
%
\bibitem{tHooVescfunc} G. 't Hooft and M. Veltman, Scalar one-loop integrals, Nucl. Phys. {\bf B153}, 365 (1979).
%
\bibitem{Jegerlehner} F. Jegerlehner, Facts of life with $\gamma_5$, Eur. Phys. J. C {\bf 18}, 673 (2001). 
%
\bibitem{Adler} S. L. Adler, Axial-vector vertex in spinor electrodynamics, Phys. Rev. {\bf 177}, 2426 (1969).
%
\bibitem{BeJa} J. S. Bell and R. Jackiw, A PCAC puzzle: $\pi^0\to\gamma\gamma$ in the $\sigma$-model, Nuovo Cimento A {\bf 60}, 47 (1969).
%
\bibitem{Fujikawa} K. Fujikawa, Comment on Chiral and Conformal Anomalies, Phys. Rev. Lett. {\bf 44}, 1733 (1980).
%
\bibitem{BrMa} P. Breitenlohner and D. Maison, Dimensionally renormalized Green's functions for theories with massless particles. I, Commun. Math. Phys. {\bf 52}, 39 (1977).
%
\bibitem{CFH} M. Chanowitz, M. Furman, and I. Hinchliffe, The axial current in dimensional regularization, Nucl. Phys. {\bf B159}, 225 (1979).
%
\bibitem{AoTo} S. Aoyama and M. Tonin, The dimensional regularization of chiral gauge theories and generalized Slavnov-Taylor identities, Nucl. Phys. {\bf B179}, 293 (1981).
%
\bibitem{Bonneau} G. Bonneau, Some fundamental but elementary facts of renormalization and regularization: A critical review of the eighties, Int. J. Geom. Methods Mod. Phys. {\bf 05}, 3831 (1990). 
%
\bibitem{NeVe} W. L. van Neerven and J. A. M. Vermaseren, Large loop integrals, Phys. Lett. {\bf 137B}, 241 (1984).
%
\bibitem{BuPi} C. P. Burgess and A. Pilaftsis, Anomalous vector-boson couplings in Majorana neutrino models, Phys. Lett {\bf B333}, 427 (1994). 
%
\bibitem{SuperKamiokandemixing} K. Abe {\it et al}. (Super-Kamiokande Collaboration), Solar neutrino measurements in Super-Kamiokande-IV, Phys. Rev. D {\bf 94}, 052010 (2016).
%
\bibitem{T2K2} K. Abe {\it et al}. (The T2K Collaboration), Constraint on the matter-antimatter symmetry-violating phase in neutrino oscillations, Nature (London) {\bf 580}, 339 (2020).
%
\bibitem {MinosPlusmixing} P. Adamson {\it et al}. (MINOS+ Collaboration), Precision Constraints for Three-Flavor Neutrino Oscillations from the Full MINOS and MINOS Dataset, Phys. Rev. Lett. {\bf 125}, 131802 (2020).
%
\bibitem{Novamixing} M. A. Acero {\it et al}. (NOvA Collaboration), First Measurement of Neutrino Oscillation Parameters using Neutrinos and Antineutrinos by NOvA, Phys. Rev. Lett. {\bf 123}, 151803 (2019).
%
\bibitem{IceCubemixing} M. G. Aartsen {\it et al}. (IceCube Collaboration), Measurement of Atmospheric Neutrino Oscillations at 6-56 GeV with IceCube DeepCore, Phys. Rev. Lett. {\bf 120}, 071801 (2018).
%
\bibitem{SuperKamiokandeanothermixing} K. Abe {\it et al}. (Super-Kamiokande Collaboration), Atmospheric neutrino oscillation analysis with external constraints in Super-Kamiokande I-IV, Phys. Rev. D {\bf 97}, 072001 (2018).
%
\bibitem{DoubleChoozmixing} H. de Kerret {\it et al}. (The Double Chooz Collaboration), Double Chooz $\theta_{13}$ measurement via total neutron capture detection, Nature Phys. {\bf 16}, 558 (2020).
%
\bibitem{Renomixing1} C. D. Shin {\it et al}. (The RENO Collaboration), Observation of reactor antineutrino disappearance using delayed neutron capture on hydrogen at RENO, J. High Energy Phys. {\bf 04}, 029 (2020).
%
\bibitem{Renomixing2} G. Bak {\it et al}. (RENO Collaboration), Measurement of Reactor Antineutrino Oscillation Amplitude and Frequency at RENO, Phys. Rev. Lett. {\bf 121}, 201801 (2018).
%
\bibitem{DayaBaymixing1} D. Adey {\it et al}. (The Daya Bay Collaboration), Measurement of the Electron Antineutrino Oscillation with 1958 Days of Operation at Daya Bay, Phys. Rev. Lett. {\bf 121}, 241805 (2018).
%
\bibitem{DayaBaymixing2} F. P. An {\it et al}. (Daya Bay Collaboration), New measurement of $\theta_{13}$ via neutron capture on hydrogen at Daya Bay, Phys. Rev. D {\bf 93}, 072011 (2016).
%
\bibitem{T2K1} K. Abe {\it et al}. (T2K Collaboration), Observation of Electron Neutrino Appearance in a Muon Neutrino Beam, Phys. Rev. Lett. {\bf 112}, 061802 (2014).
%
\bibitem{CMShnl} A. M. Sirunyan {\it et al}. (CMS Collaboration), Search for heavy neutral leptons in events with three charged leptons in proton-proton collisions at $\sqrt{s}=13\,{\rm TeV}$, Phys. Rev. Lett. {\bf 120}, 221801 (2018).
%
\bibitem{FGLY} E. Fern\'andez-Mart\'inez, M. B. Gavela, J. L\'opez-Pav\'on, and O. Yasuda, CP-violation from non-unitary leptonic mixing, Phys. Lett. B {\bf 649}, 427 (2007).
%
\bibitem{FHL} E. Fern\'andez-Mart\'inez, J. Hern\'andez-Garc\'ia, and J. L\'opez-Pav\'on, Global constraints on heavy neutrino mixing, JHEP {\bf 08}, 033 (2016).
%
\bibitem{BFHLMN} M. Blennow, E. Fern\'andez-Mart\'inez, J. Hern\'andez-Garc\'ia, J. L\'opez-Pav\'on, X. Marcano, and D. Naredo-Tuero, Bounds on lepton non-unitarity and heavy neutrino mixing, JHEP {\bf 08}, 030 (2023). 
%
\bibitem{D0onWWZ} V. M. Abazov {\it et al}. (D0 Collaboration), Limits on anomalous trilinear gauge boson couplings from $WW$, $WZ$, and $W\gamma$ production in $p\bar{p}$ collisions at $\sqrt{s}=1.96\,{\rm TeV}$, Phys. Lett. B {\bf 718}, 451 (2012).
%
\bibitem{ATLASonWWZ}  M. Aaboud {\it et al}. (ATLAS Collaboration), Measurement of $WW/WZ\to l\nu qq'$ production with the hadronically decaying boson reconstructed as one or two jets in $pp$ collisions at $\sqrt{s}=8\,{\rm TeV}$ with ATLAS, and constraints on anomalous gauge couplings, Eur. Phys. J. C {\bf 77}, 563 (2017).
%
\bibitem{CMSonWWZ} A. M. Sirunyan {\it et al}. (The CMS Collaboration), Search for anomalous couplings in boosted $WW/WZ\to l\nu q\bar{q}$ production in proton-proton collisions at $\sqrt{s}=8\,{\rm TeV}$, Phys. Lett.
B {\bf 772}. 21 (2017).
%
\bibitem{CMSonWWZimproved} A. M. Sinrunyan {\it et al}. (CMS Collaboration), Measurement
of the $W\gamma$ Production Cross Section in Proton-Proton Collisions at $\sqrt{s}=13\,{\rm TeV}$ and Constraints on Effective Field Theory Coefficients, Phys. Rev. Lett. {\bf 126}, 252002 (2021).
%
\bibitem{LEPsuperbound} S. Schael {\it et al}. (The ALEPH Collaboration), Electroweak measurements in electron-positron collisions at $W$-boson-pair energies at LEP, Phys. Rep. {\bf 532}, 119 (2013); J. Abdallah {\it et al}. (The DELPHI Collaboration), Phys. Rep. {\bf 532}, 119 (2013); P. Achard {\it et al}. (The L3 Collaboration), Phys. Rep. {\bf 532}, 119 (2013); G. Abbiendi {\it et al}. (The OPAL Collaboration), Phys. Rep. {\bf 532}, 119 (2013).
%
\bibitem{DelphiCPodd} J. Abdallah {\it et al}. (DELPHI Collaboration), Study of $W$-boson polarisations and triple gauge boson couplings in the reaction $e^+e^-\to W^+W^-$ at LEP 2, Eur. Phys. J. C {\bf 54}, 345 (2008).
%
\bibitem{CEPConWWZ} L. Bian, J. Shu, and Y. Zhang, Prospects for triple gauge coupling measurements at future lepton colliders and the 14 TeV LHC, J. High Energy Phys. {\bf 09}, 206 (2015).
%
\bibitem{AKM} A. Arhrib, J. -L. Kneur, and G. Moultaka, MSSM radiative contributions to the $WW\gamma$ and $WWZ$ form factors, Phys. Lett.  {\bf 376}, 127 (1996).
%
\bibitem{MTTR} J. Monta\~no, G. Tavares-Velasco, J.J. Toscano, and F. Ram\'irez-Zavaleta, $SU_L(3)\times U_X(1)$-invariant description of the bilepton contributions to the $WWV$ vertex in the minimal 331 model, Phys. Rev. D {\bf 72}, 055023 (2005).
%
\bibitem{PiPl} F. Pisano and V. Pleitez, ${\rm SU}(3)\otimes{\rm U}(1)$ model for electro-
weak interactions, Phys. Rev. D {\bf 46}, 410 (1992).
%
\bibitem{Frampton} P.H. Frampton, Chiral Dilepton Model and the Flavor Question, Phys. Rev. Lett. {\bf 69}, 2889 (1992).
%
\bibitem{RTT} F. Ram\'irez-Zavaleta, G. Tavares-Velasco, and J. J. Toscano, Bilepton effects on the $WWV$ vertex in the 331 model with right-handed neutrinos via a $SU_L(2)\times U_Y(1)$ covariant quantization scheme, Phys. Rev. D {\bf 75}, 075008 (2007).
%
\bibitem{ACD} T. Appelquist, H. -C. Cheng, and B. A. Dobrescu, Phys. Rev. D {\bf 64}, 035002 (2001).
%
\bibitem{FMNRT} A. Flores-Tlalpa, J. Monta\~no, H. Novales-S\'anchez, F. Ram\'irez-Zavaleta, and J. J. Toscano, One-loop effects of extra dimensions on the $WW\gamma$ and $WWZ$ vertices, Phys. Rev. D {\bf 83}, 016011 (2011).
%
\bibitem{LMMNTT} M. A. L\'opez-Osorio, E. Mart\'inez-Pascual, J. Monta\~no, H. Novales-S\'anchez, J. J. Toscano, and E. S. Tututi, Trilinear gauge boson couplings in the standard model with one universal extra dimension, Phys. Rev. D {\bf 88}, 016010 (2013).
%
\bibitem{GeMa} H. Georgi and M. Machacek, Doubly charged Higgs bosons, Nucl. Phys. {\bf B262}, 463 (1985).
%
\bibitem{UHT} M. A. Arroyo-Ure\~na, G. Hern\'andez-Tom\'e, and G. Tavares-Velasco, $WWV$ ($V=\gamma,Z$) vertex in the Georgi-Machacek model, Phys. Rev. D {\bf 94}, 095006 (2016).
%
\bibitem{Sakharov} A. D. Sakharov, Violation of CP invariance, $C$ asymmetry, and baryon asymmetry of the universe, Pis'ma Zh. Eksp. Teor. Fiz. {\bf 5}, 32 (1967) JETP Lett. {\bf 5}, 24 (1967); Usp. Fiz. Nauk {\bf 161}, 61 (1991) Sov. Phys. Usp. {\bf 34}, 392 (1991).
%
\bibitem{KoMa} M. Kobayashi and T. Maskawa, CP-violation in the renormalizable theory of weak interaction, Prog. Theor. Phys. {\bf 49}, 652 (1973).
%
\bibitem{CPoddSusy} M. Kitahara, M. Marui, N. Oshimo, T. Saito, and A. Sugamoto, $CP$-odd anomalous $W$-boson couplings from supersymmetry, Eur. Phys. J. C {\bf 4 }, 661 (1998).
%
\bibitem{AMOSS} E. Asakawa, M. Marui, N. Oshimo, T. Saito, and A. Sugamoto, $CP$-odd $WWZ$ couplings induced by vector-like quarks, Eur. Phys. J. C {\bf 10}, 327 (1999).
%
\bibitem{AppWu} T. Appelquist and G. -H. Wu, Electroweak chiral Lagrangian and $CP$-violating effects in technicolor theories, Phys. Rev. D {\bf 51}, 240 (1994).
%
\bibitem{DiNa} M. Diehl and O. Nachtmann, Optimal observables for the measurement of three gauge boson couplings in $e^+e^-\to W^+W^-$, Z. Phys. C {\bf 62}, 397 (1994).
%
\bibitem{Takagi} T. Takagi, On an algebraic problem related to an analytic theorem of Carath\'eodory and Fej\'er and on an allied theorem of Landau, Jpn. J. Math. {\bf 1}, 83 (1925). 
%
\bibitem{KPS} J. G. K\"orner, A. Pilaftsis, and K. Schilcher, Leptonic CP asymmetries in flavor-changing $H^0$ decays, Phys. Rev. D {\bf 47}, 1080 (1993).
%
\bibitem{DePi} P. S. B. Dev and A. Pilaftsis, Minimal radiative neutrino mass mechanism for inverse seesaw models, Phys. Rev. D {\bf 86}, 113001 (2012).
%
\end{thebibliography}
\end{document}